\pdfoutput=1
\documentclass[cernpreprint, texlive=2016, UKenglish, texmf]{atlasdoc}
 
\usepackage[subcaption, rotating, longtable, block=space]{atlaspackage}
\usepackage[italic]{hepnicenames}
\usepackage{multirow}
\usepackage{adjustbox}
\usepackage{enumerate}
\usepackage{dcolumn}
\usepackage{floatrow}
\newfloatcommand{capbtabbox}{table}[][\FBwidth]
\usepackage{blindtext}
\usepackage{bm}
\usepackage{atlasbiblatex}
\usepackage{atlascontribute}
\usepackage[hepparticle, hepprocess, xref=true, unit=false]{atlasphysics}
 
\graphicspath{{logos/}{figures/}}
 
\usepackage{ANA-TOPQ-2018-10-PAPER-defs}

\AtlasTitle{Measurement of the polarisation of single top~quarks and antiquarks produced in the $t$-channel at $\sqrt{s}=13$~TeV and bounds on the \textit{tWb} dipole operator from the ATLAS experiment}
 
\AtlasAbstract{
A simultaneous measurement of the three components of the
top-quark and top-antiquark polarisation vectors in $t$-channel single-top-quark
production is presented. This analysis is based on data from proton--proton collisions at
a centre-of-mass energy of 13 TeV corresponding to an
integrated luminosity of 139 fb$^{-1}$, collected with the ATLAS detector at
the LHC. Selected events contain exactly one isolated electron or muon, large
missing transverse momentum and exactly two jets, one being $b$-tagged.
Stringent selection requirements are applied to discriminate $t$-channel
single-top-quark events from the background contributions.
The top-quark and top-antiquark polarisation vectors are measured from the distributions of the direction cosines of the charged-lepton momentum in the top-quark rest frame.
The three components of
the polarisation vector for the selected top-quark event sample are
$P_{x'} =0.01\pm0.18$, $P_{y'} = -0.029\pm0.027$, $P_{z'} =0.91\pm0.10$
and for the top-antiquark event sample they are  $P_{x'} = -0.02\pm0.20$, $P_{y'} = -0.007\pm0.051$,
$P_{z'} = -0.79\pm 0.16$.
Normalised differential cross-sections corrected to a fiducial region
at the stable-particle level are presented as a function of
the charged-lepton angles for top-quark and top-antiquark
events inclusively and separately.
These measurements are in
agreement with Standard Model predictions.
The angular differential cross-sections are used to derive bounds on the complex Wilson coefficient
of the dimension-six $\mathcal{O}_{tW}$ operator in
the framework of an effective field theory.
The obtained bounds are $C_{tW} \in\ [-0.9,1.4]$ and $C_{itW} \in [-0.8,0.2]$, both at 95\% confidence level.}
\author{The ATLAS Collaboration}
 
\PreprintIdNumber{CERN-EP-2021-263}

\AtlasJournalRef{\JHEP 11 (2022) 040}
\AtlasDOI{DOI:10.1007/JHEP11(2022)040}

\hypersetup{pdftitle={ATLAS document},pdfauthor={The ATLAS Collaboration}}
 
\begin{document}
\maketitle

\section{Introduction}
\label{sec:intro}
\begin{figure}[!htpb]
\begin{center}
\subcaptionbox{}{\includegraphics[width=0.45\textwidth]{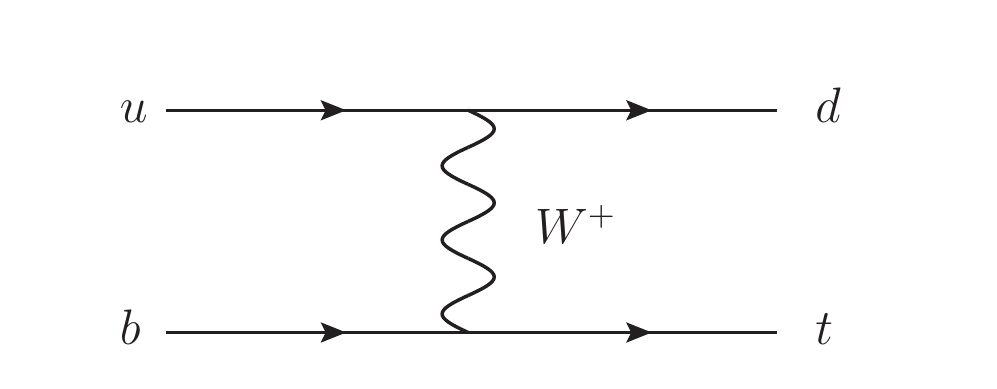}}
\subcaptionbox{}{\includegraphics[width=0.45\textwidth]{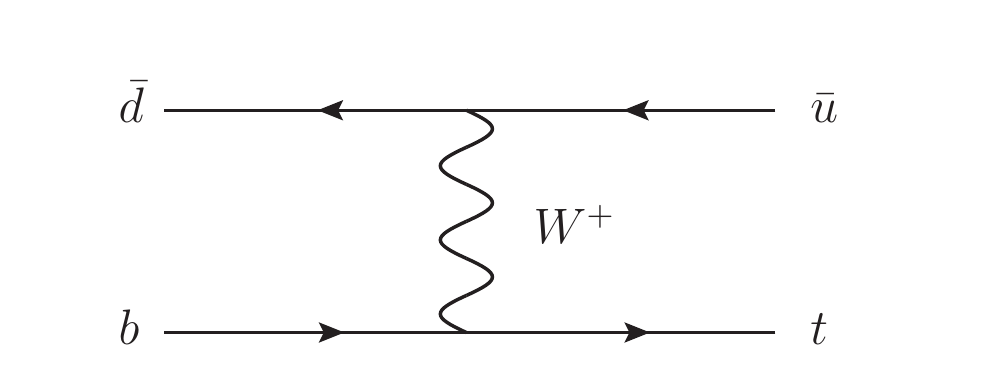}}
\subcaptionbox{}{\includegraphics[width=0.45\textwidth]{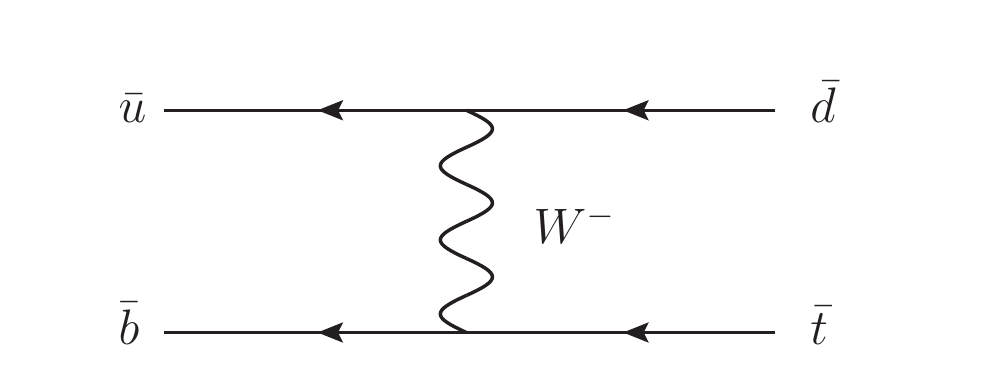}}
\subcaptionbox{}{\includegraphics[width=0.45\textwidth]{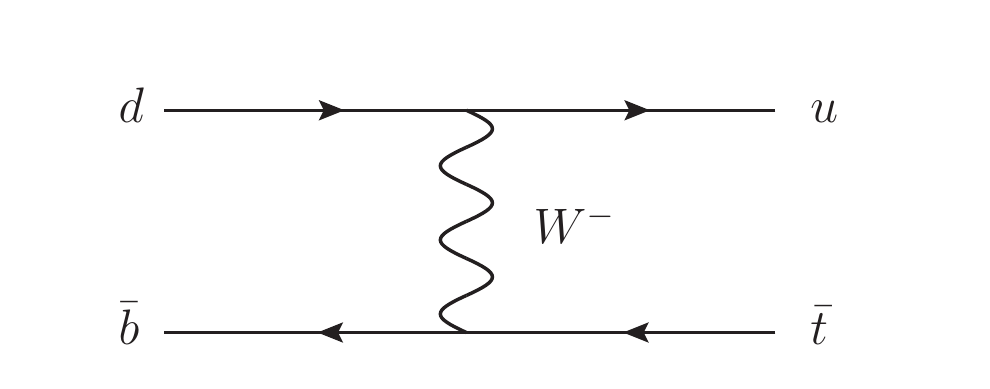}}
\caption{Feynman diagrams for the processes contributing to \tch\ single-top-quark production at LO, in the five-flavour scheme.
In the dominant subprocess, an up- or down-type quark from one of the
colliding protons interacts with a bottom quark or antiquark from the other proton
by exchanging a virtual $W$ boson to produce a (a) top quark or
(d) top antiquark. In the subdominant subprocess, a down- or up-type antiquark from one of the
colliding protons interacts with a bottom quark or antiquark from another proton
by exchanging a virtual $W$ boson to produce a (b) top quark or
(c) top antiquark.}
\label{fig:productionSubprocesses}
\end{center}
\end{figure}

Single-top-quark production through the electroweak (EW) charged current at hadron colliders proceeds mostly, according to the Standard Model (SM)
prediction, via three modes that can be defined at leading order (LO)
in quantum chromodynamics (QCD): the exchange of a
virtual $W$ boson in either the $t$- or \sch, and the associated
production of a top quark and a $W$ boson (named \tWch).
At the LHC, in proton--proton ($pp$) collision data, the \tch is the dominant process
and the subject of the measurements presented in this paper.
In the \tch\ process, a light-flavour quark $q$ from one of the colliding protons
interacts with a $b$-quark, which can be considered as being emitted directly
from the other colliding proton (five-flavour scheme or 5FS)
or as originating from gluon splitting (four-flavour scheme or 4FS). The incoming light-flavour quark
exchanges a space-like virtual $W$ boson, producing a top quark $t$
and a recoiling light-flavour quark $q'$, called the \emph{spectator quark}.
Two subprocesses contribute to the \tch\ process in the production of either single top quarks ($t$) or
single top antiquarks ($\bar{t}$) at LO. The dominant subprocess
is the scattering of the incoming up-type (down-type) quark from a
bottom quark (antiquark), to produce a down-type (up-type) spectator quark
and a top quark (antiquark), as illustrated in
Figures~\ref{fig:productionSubprocesses}(a) and~\ref{fig:productionSubprocesses}(d).
The subdominant subprocess is the scattering
of a down-type (up-type) antiquark from a bottom quark (antiquark),
to produce an up-type (down-type) spectator antiquark and a top quark
(antiquark), as illustrated in Figures~\ref{fig:productionSubprocesses}(b)
and~\ref{fig:productionSubprocesses}(c).
The production cross-section of single top quarks
is about twice as large as that of single top antiquarks.
 
The QCD $pp \rightarrow \ttbar$ process produces unpolarised top
quarks because of parity conservation in QCD~\cite{Kane:1978nd,
Devoto:1979jm, PhysRevD.45.124, Mahlon:1995zn},
while single-top-quark production yields a large sample of highly
polarised top quarks and top antiquarks.
In the \tch\ at LO, as a
consequence of the vector minus axial-vector $(V-A)$ form of the \Wtb\ vertex,
single top quarks are produced with their spin completely aligned
along the direction of the down-type quarks~\cite{Mahlon:1996pn, Mahlon:1999gz}.
For single top quarks produced by the dominant subprocess,
this direction is the spectator-quark direction, while for the
subdominant subprocess
it is the direction of the incoming down-type antiquark. For single top-antiquark production, the spin aligns
in the direction opposite to that of the incoming down-type quark in
the dominant subprocess, and opposite to that of the spectator antiquark in the subdominant process.
Thus, the degree of polarisation for a sample of single-top-quark or single-top-antiquark events depends on the mix of dominant and subdominant production processes and the relative alignment between the beam line and spectator-quark directions, averaged over the sample selected. The orientation of the polarisation vector is in the production plane.
In all cases the direction is relative
to the rest frame of the top quark or antiquark.
 
In the \tch\ production process, the spectator-quark direction lies close to the beam direction.
The predominant $ub\rightarrow dt$ process (or
$ug\rightarrow dt\bar{b}$ in the 4FS), as well as the similarity
of the two aforementioned directions, produces very high
(though not 100\%) top-quark polarisation along the direction of the spectator
quark when both the dominant and subdominant processes are considered.
For top antiquarks, the two subprocesses contribute to different degrees,
but also lead to nearly 100\% polarisation, this time in the direction
opposite to that of the spectator quark. At LO in the SM, the expected values of the polarisations of top quarks and top antiquarks along the direction of the
spectator quark are 0.90 and $-$0.86, respectively, computed in the 4FS
as detailed in Ref.~\cite{Aguilar-Saavedra:2014eqa}. However, effects beyond LO as well as acceptance requirements, which  have a large effect on the polarisation, are not accounted for in that calculation.
A calculation at next-to-next-to-leading order (NNLO), based on Ref.~\cite{Berger:2017zof}, predicts top-quark polarisation along the direction of the spectator quark
of $0.965 \pm{0.003} ~\text{(scale)}{}^{+0.003}_{-0.004} ~\text{(PDF\,+\,\alphas)}$ for top quarks and
$-0.957 {}^{-0.003}_{+0.012}~\text{(scale)}{}^{+0.004}_{-0.002}~\text{(PDF\,+\,\alphas)}$ for top antiquarks. These values are
obtained from a parton-level calculation at fixed order using the \CT[18nnlo]~\cite{Hou:2019efy} set of parton
distribution functions (PDFs) with the renormalisation ($\muR$) and factorisation ($\muF$)
scales set to half of the top-quark mass ($\mt$). This calculation considers stable
single top quarks produced in the \tch\ from $pp$ collisions at
\cmenergyThirteen, in a region with exactly one light-flavour jet, where jets are clustered with the
\antikt\ algorithm~\cite{Cacciari:2008gp} with a radius parameter of 0.4, and they are required to have
$\pt > \SI{30}{\GeV}$ and $|\eta|<4.5$.
Additional acceptance criteria for the light-flavour jet, implementing
the acceptance requirement defined in \Equation{eq:trap} and described
in \Section{sec:event-selection},
were also applied in calculating the prediction.
The effect of the scale uncertainty is calculated by varying the renormalisation and
factorisation scales
by factors of 2.0 and 0.5 from their
central value.
The effect of the \alphas\ uncertainty is calculated by varying $\alphas(\mZ)$ by 0.001
from its central value. The PDF uncertainty is calculated from the
68\% confidence level (CL) eigenvectors in the \CT[18nnlo] PDF sets
at the most similar \alphas\ value.
The effects of the \alphas\ and PDF uncertainties are summed in
quadrature.
Similar calculations were
previously carried out for $pp$ collisions at LO at \cmenergyForteen~\cite{Mahlon:1999gz},
at next-to-leading order (NLO) at \cmenergySeven~\cite{Schwienhorst:2010je},
and in proton--antiproton collisions at LO at Tevatron
energies~\cite{Mahlon:1996pn}. A general trend is that typical
acceptance requirements increase the top-quark polarisation, while
effects beyond LO reduce it.
NLO EW corrections to the \Wtb\ vertex are calculated in
Refs.~\cite{Do_2003,Gonz_lez_Sprinberg_2011,Gonz_lez_Sprinberg_2015,Arhrib_2016,Fischer_2018},
and Ref.~\cite{Arhrib_2016} explicitly calculates the effect on
top-quark polarisation.  While that calculation is done for the full
phase space of single-top-quark production, rather than the restricted
space of the calculation described in this paper, the EW corrections are smaller than those from QCD.
 
This paper reports a direct measurement of the top-quark and
top-antiquark polarisation vectors from a template fit to the joint
distributions of the direction cosines of the charged-lepton momentum
in the top-quark rest frame. Each component of the polarisation vector
is thereby measured without any assumption about the other two
components. In addition, normalised differential cross-sections
corrected to a fiducial region at particle level are presented as a
function of the charged-lepton angles for top-quark and top-antiquark
events separately and inclusively, so that they can be combined with other experimental inputs,
to derive bounds on complex Wilson coefficients in the framework
of an effective field theory (EFT). In this paper, the inclusive
measurements are used to derive bounds on the complex Wilson
coefficient of the dimension-six $\mathcal{O}_{tW}$ operator.
 
\subsection{Decay angles from polarised top quarks and top antiquarks}
\label{sec:axes_definition}
 
\begin{figure}[!htbp]
\begin{center}
\includegraphics[width=0.52\textwidth]{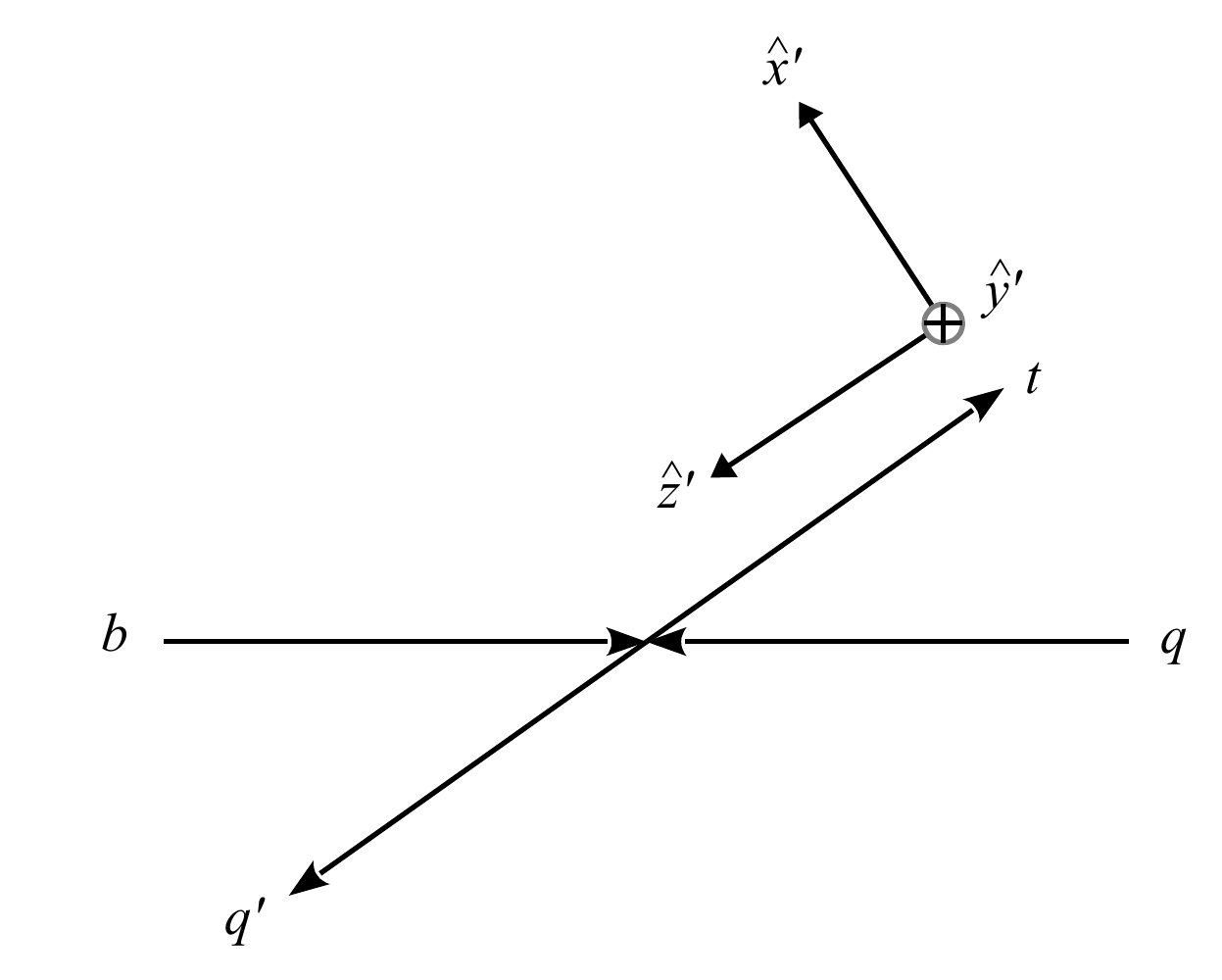}
\caption{Diagram illustrating the three orthogonal directions $\hat{x}'$,
$\hat{y}'$ and $\hat{z}'$ used in this analysis, as seen in the
zero-momentum frame of the initial-state quarks. The $\hat{z}'$ direction is that of the
spectator quark in the top-quark rest frame. The $\hat{x}'$
direction lies in the production plane, while the $\hat{y}'$
direction is perpendicular to the production plane.}
\label{fig:axes_definition}
\end{center}
\end{figure}

This analysis exploits the $t\rightarrow Wb \rightarrow b\ell^+\nu$ decay mode of the top quark, as well as the charge-conjugate decay mode of the top antiquark.  The lepton $\ell^{\pm}$ can be either an electron or a muon. In the decay, three orthogonal directions may be defined~\cite{Aguilar-Saavedra:2014eqa}. These serve to express the spin vector of the top quark. As illustrated in \Figure{fig:axes_definition}, the $\hat{z}'$ direction is the direction of the momentum of the
spectator quark, $\vec{p}_{q^\prime}$, in the top-quark reference frame. The $\hat{y}'$
direction is taken along $\hat{z}'\times\hat{p}_q$, where
$\hat{p}_q$ is the direction of the incoming light-flavour quark, in the
top-quark reference frame.
Then, the $\hat{x}'$ direction lies in the
production plane, orthogonal to $\hat{y}'$ and $\hat{z}'$, such that
$\{x',y', z'\}$ form a right-handed coordinate system:
$\hat{z}' =\vec{p}_{q^\prime } / |\vec{p}_{q^\prime}|$, $\hat{y}'= ( \hat{z}'\times\vec{p}_q )/ |\hat{z}'\times\vec{p}_q|$ and
$\hat{x}'=\hat{y}'\times\hat{z}'$.
 
The polarisation vector $\vec{P}$ is defined in this coordinate system; it satisfies $|\vec{P}| \leq 1$, equality holding if and only if the top quarks are produced in a pure quantum mechanical ensemble with respect to spin.
Information about the spin of the top quark is transferred
to the decay products, and therefore can be extracted
from their angular distributions. This was already exploited in previous
analyses~\cite{TOPQ-2015-13, TOPQ-2013-12,TOPQ-2014-17, TOPQ-2015-11, CMS-TOP-17-023} which measure
spin observables in single-top-quark and/or \ttbar\ events using LHC data. Previous
measurement of \tch\ single-top-quark polarisation at \cmenergyEight
from ATLAS~\cite{TOPQ-2015-11} set a limit $\left |\Polz \right | >
0.72$ (at 95\% CL) at parton level, where an
average is taken over top quarks and top antiquarks.
The spin asymmetry $A_\ell=(\Polz \alpha_{\ell})/2$, where $\alpha_{\ell}$ is the analysing power of the charged lepton ($\ell$) in
the top-quark decay, was determined to be $0.49 \pm 0.06$ by ATLAS~\cite{TOPQ-2014-17}
and $0.26 \pm 0.11$ by CMS~\cite{CMS-TOP-13-001} at \cmenergyEight\
and measured to be $0.440 \pm 0.070$ by CMS at
\cmenergyThirteen~\cite{CMS-TOP-17-023}.
In all these cases, this was done
at parton level, and an average over top quarks and antiquarks was
also taken.
 
As shown in Refs.~\cite{Kane:1978nd,
Devoto:1979jm, PhysRevD.45.124, Mahlon:1995zn}, the charged lepton is
the most sensitive probe of the top-quark spin, with analysing
power close to 1; for that reason the analysis is based upon angular
distributions of the charged lepton from the top-quark decay.
Angular distributions of single top quarks are
discussed in Ref.~\cite{Aguilar-Saavedra:2017wpl}, where the four-dimensional fully differential decay distribution for the
top-(anti)quark decays with polarisation $\vec{P}$ is presented. Equation~(11) in that reference is used as the basis for a custom decay model, valid at LO, which is interfaced with the LO \PROTOS~\cite{AguilarSaavedra:2008gt}
event generator, in order to generate simulated event samples
containing pure ensembles of events fully polarised along the $x'$,
$y'$, or $z'$ direction.
Calculations of top-quark angular decay distributions at NLO and NNLO~\cite{Fischer:2001gp,Czarnecki:2010gb} indicate that higher-order effects are small compared to the precision of this analysis.
A fitting function describing angular distributions of event samples with arbitrary polarisation is constructed by taking linear combinations of templates derived from simulated pure ensembles with $x'$, $y'$, or $z'$ polarisation at reconstruction level.
The fitting function is then used in a template fit
to data, in which the three components of polarisation,
$\{\Polx,\Poly,\Polz\}$ are allowed
to float without imposing the $|\vec{P}| \leq 1$ constraint.  This is the strategy followed in this
analysis to measure all components of the polarisation vector $\vec{P}$ for top-quark and top-antiquark events separately.
Details of the fit model are given in \Section{sec:template-fit}.
 
In the coordinate system previously defined, and considering $\theta_{\ell i}$ as the polar
angle of the charged-lepton momentum with respect to the $i^\mathrm{th}$ axis
($i = x', y', z'$), one can obtain the differential angular distributions
associated with the three different polarisation components $\{\Polx,\Poly,\Polz\}$.
These differential angular distributions are distorted by the inefficiencies and acceptance of the detector, smeared by the reconstruction procedures, and sculpted by the
event selection criteria. In this paper, the detector effects are
unfolded and the measured normalised differential angular distributions
are compared directly with theoretical predictions within a
fiducial region as detailed in \Section{sec:unfolding-method}.
 
\subsection{Sensitivity of polarised top-quark decay angles to effective field theory operators}
\label{sec:EFT_intro}
 
Measurements of polarisation observables in \tch\ single-top-quark production are sensitive to new physics phenomena
affecting the \Wtb\ vertex.
In the framework of SM EFT~\cite{BUCHMULLER1986621}, the SM Lagrangian ($\mathcal{L}_{\text{SM}}$) is augmented with higher-dimensional operators invariant under the SM gauge symmetry.
Since dimension-five operators do not contribute to top-quark
production or decay~\cite{Zhang:2010dr}, the leading contributions from higher-dimensional operators are from dimension-six operators ($\mathcal{O}^{[6]}_k$). This analysis is therefore limited to those terms, with corresponding Wilson coefficients ($C_k$) scaled by $1/\Lambda^2$:
 
\begin{equation*}
\mathcal{L}_{\text{EFT}} = \mathcal{L}_{\text{SM}} +  \sum_k \frac{C_k}{\Lambda^2} \mathcal{O}^{[6]}_k  + ... \,,
\end{equation*}
 
where $\mathcal{L}_{\text{EFT}}$ is the effective Lagrangian, $\Lambda$ is
the scale of new physics chosen such that higher-dimensional operators
are sufficiently suppressed by higher powers of $\Lambda$, and where $k$ runs over all dimension-six operators.
There are only three dimension-six
effective operators that contribute at order $1/\Lambda^2$ at tree level in
the \tch\ production of the single top quark
as presented in Ref.~\cite{deBeurs:2018pvs}.
In the so-called `Warsaw' basis, the three operators are $\mathcal{O}_{\varphi
q}$, $\mathcal{O}_{qq}$ and $\mathcal{O}_{tW}$ as described in
Ref.~\cite{Grzadkowski:2010es}, where $\varphi$ stands for the Higgs field.
Other effective operators such as $\mathcal{O}_{bW}$ and $\mathcal{O}_{\varphi tb}$ contribute but are suppressed by a factor
$1/\Lambda^4$ or by the small value of the bottom-quark mass relative to that of the top quark. Within the present limits set on
these other operators~\cite{Brivio:2019ius,Hartland:2019bjb,Durieux:2019rbz,TOPQ-2015-11,ethier2021combined,Deliot2018,TOPQ-2018-02}, their effect on the polarisation of the top quark is expected to be insignificant compared to the current precision of the measurement.
 
The operator $\mathcal{O}_{\varphi q}$ affects the signal production cross-section and
has no effect on normalised distributions. The four-fermion operator,
$\mathcal{O}_{qq}$, has a negligible effect on angular
distributions~\cite{deBeurs:2018pvs} and can be ignored.
Only the \OtW\ operator with its complex coefficient has an effect on the polarisation of the top quark.
The real and imaginary parts of its coefficient are indicated in this
work by \ctW\ and \citW, respectively.
The coefficient \ctW\ mostly affects $\Polx$, whereas \citW\ affects $\Poly$.
Since the SM prediction for $\Polz$ is already near the maximum allowed value, it is less sensitive to changes in \ctW\ or \citW.
A non-zero value for \citW\ could be a hint of non-SM CP violation in the
\Wtb\ vertex, making it especially interesting with regard to the matter--antimatter asymmetry present in nature.
 
This formalism is used to perform an interpretation of the unfolded
normalised differential angular distributions
presented in \Section{sec:axes_definition} in an EFT context,
providing limits on the real and imaginary parts of the $\mathcal{O}_{tW}$ dipole operator.
 
This paper is organised as follows.
\Section{sec:data_mc_samples} describes the data
samples as well as the simulated event samples used to predict
properties of the \tch\ signal and the background
processes. \Section{sec:event-reconstruction} describes the object and
event reconstruction for the identification of signal events,
while \Section{sec:event-selection} presents the criteria used to define
the signal region and the control regions. The
procedures for modelling background processes are described
in \Section{sec:background-estimation}. The event yields comparing the
predictions and the observed data are also shown in this
section. \Section{sec:systematics} quantifies the systematic
uncertainties in this measurement. \Section{sec:template-fit} describes
the measurement of the polarisation vector of an ensemble of top quarks or
antiquarks. The particle-level differential measurement of the normalised angular
distributions is described in \Section{sec:unfolding-method}, while the way these angular
differential cross-section distributions are used to set limits on some
EFT coefficients is presented in \Section{sec:eft_interpretation}.
The conclusions are given in \Section{sec:conclusions}.

\section{Data and simulated event samples}
\label{sec:data_mc_samples}
 
The ATLAS detector\footnote{ATLAS uses a right-handed coordinate system with its origin at the nominal interaction point (IP) in the centre of the detector and the $z$-axis along the beam pipe. The $x$-axis points from the IP to the centre of the LHC ring, and the $y$-axis points upward. Cylindrical coordinates $(r,\phi)$ are used in the transverse plane, $\phi$ being the azimuthal angle around the $z$-axis. The pseudorapidity is defined in terms of the polar angle $\theta$ as $\eta=-\ln\tan(\theta/2)$.}~\cite{PERF-2007-01} at the LHC covers nearly the entire solid angle around the collision point. It consists of an inner tracking detector surrounded by a thin superconducting solenoid producing a \SI{2}{\tesla} axial magnetic field, electromagnetic and hadronic calorimeters, and an external muon spectrometer incorporating three large toroidal magnet assemblies.
The analysis is performed using $pp$ collision data collected at a centre-of-mass energy of \SI{13}{\TeV} from \datafirstyear\ to \datalastyear. Stringent detector and data quality requirements were applied~\cite{DAPR-2018-01}, resulting in a data sample corresponding to a total integrated luminosity of \lumi~\cite{ATLAS-CONF-2019-021, LUCID2}.
The events were selected by single-lepton\footnote{Henceforth, `lepton' indicates an electron or muon.} triggers~\cite{TRIG-2018-05, TRIG-2018-01}, imposing low thresholds on the transverse energy ($\ET$) of electrons and the transverse momentum ($\pT$) of muons, in addition to isolation requirements, or imposing a looser identification criterion and higher thresholds with no isolation requirement. The lowest thresholds varied from 20 to \SI{26}{\GeV}  depending on the lepton flavour and the data-taking period.
 
Samples of simulated events were produced using different Monte Carlo (MC) event generators including parton shower (PS) and hadronisation models.
The effect of multiple interactions in the same and neighbouring bunch crossings (pile-up)
was modelled by overlaying the hard-scattering event with simulated minimum-bias events
generated with \PYTHIA[8.186]~\cite{Sjostrand:2007gs, Sjostrand:2014zea} using the
\NNPDF[2.3lo]~\cite{Ball:2012cx} PDF set and the A3 set of tuned parameters~\cite{ATL-PHYS-PUB-2016-017}.
Events were reweighted such that the
distribution of the average number of interactions per bunch crossing
matches that observed in data.

Single-top-quark \tch\ events were generated with the NLO \POWHEGBOX~\cite{Frederix:2012dh,Nason:2004rx,Frixione:2007vw,Alioli:2010xd} v2 generator, which provides matrix elements (MEs) at NLO in the strong coupling constant \alphas\ using the 4FS with the \NNPDF[3.0nlo]{}~nf4~\cite{Ball:2014uwa} PDF set.
The scales were set to $\muR^2 = \muF^2  = 16(\mb^2 + \pTb^2)$, where $\mb$ and $\pTb$ are the mass and \pT of the $b$-quark from the initial gluon splitting.
 
Additional samples of simulated \tch\ signal events were produced within the 4FS with the LO \PROTOS[2.3] generator~\cite{AguilarSaavedra:2008gt} using the \cteq PDF set~\cite{Pumplin:2002vw}.
The scales were set to $\muF^2 = - p^2_W$ for the light-flavour quark distribution function and $\muF^2 = \pTbbar^2 + m^2_b$ for the gluon distribution function, where $p_W$ and $\pTbbar$ are the momentum of the exchanged $W$ boson and the \pT\ of the $b$-antiquark originating from the gluon splitting, respectively~\cite{AguilarSaavedra:2008gt,Stelzer:1997ns}. In addition to a SM signal sample, six samples with $\tWb$ anomalous couplings~\cite{Boos:1999dd,Chen:2005vr}, \vlr\ and \glr, enabled in both the production and decay vertices were produced with the \PROTOS\ event generator. The ranges chosen for the anomalous coupling event samples were based on previous established limits~\cite{TOPQ-2015-11}: $\text{Re}[\gr]/\vl = \pm0.18$; $\text{Im}[\gr]/\vl = \pm0.07$; and $\vr/\vl = \pm 0.4$ and $\gl/\vl = \mp 0.32$.
For $\vl=1$ and $\Lambda=\SI{1}{\TeV}$, following Ref.~\cite{AguilarSaavedra:2008zc}, these correspond to $\ctW=\pm 2.0$, $\citW=\pm 0.8$, and $\cphitb=\pm 13$ and $\cbW=\mp 3.7$, respectively.
The \PROTOS\ generator was also modified with a special decay model allowing it to generate event samples with top quarks fully polarised along, or opposite to, the $x'$, $y'$, or $z'$ axes as defined in \Section{sec:axes_definition}, resulting in six simulated samples in total. These samples are used to produce templates for the determination of top-quark and top-antiquark polarisations.
 
Furthermore, five samples of simulated signal events with non-zero Wilson coefficients related to dimension-six effective operators were also produced. These simulation samples were generated at NLO, using the same set-up as in Ref.~\cite{deBeurs:2018pvs}, with
the \MGNLO[2.6.2]~\cite{Alwall:2014hca} generator using the \NNPDF[3.0nlo] PDF set.
The values of the non-zero Wilson coefficients in the different simulated event samples were set to: $\ctW = 2.0$; $\citW=  1.75$; $\cqq = -0.4$; $\ctW = -2.0$ and $\cqq = -0.4$; and $\ctW = 2.0$ and $\citW=  -1.75$.
The EFT operators were allowed to enter in the production vertex as well as the decay vertex, consistently including possible effects on the width of the top quark. The implementation of the NLO effective operators~\cite{Zhang_2016} makes use of the 5FS, indicating that the $b$-quark is treated as being massless and thus part of the proton.
The two scales were set to $\muR = \muF  = \mt$ in the MC generation.
 
The \tch\ signal production cross-section
was calculated at NLO~\cite{PhysRevLett.102.182003} in perturbative QCD using \HATHOR[2.1]~\cite{Aliev:2010zk,Kant:2014oha}.
For $pp$ collisions at \cmenergyThirteen, this cross-section corresponds to
$136 \pm 5$~pb and $81 \pm 4$~pb for
top-quark and top-antiquark production, respectively. The cross-section uncertainties connected with the PDFs and $\alphas$ were calculated using the PDF4LHC prescription~\cite{Butterworth:2015oua} with the \MSTW[nlo] 68\%~CL~\cite{Martin:2009iq, Martin:2009bu}, \CT[10nlo]~\cite{Gao:2013xoa} and \NNPDF[2.3nlo] PDF sets, and were added in quadrature to the effect of the scale uncertainty.
 
The production of \ttbar\ events, as well as single-top-quark events in the \Wt process and in the \sch, were modelled using the \POWHEGBOX[v2] generator, which provides MEs at NLO,
with the \NNPDF[3.0nlo] PDF set. In \ttbar\ events, the \POWHEGBOX\ \hdamp\ parameter\footnote{The \POWHEG\ \hdamp\ parameter
controls the \pT\ of the first additional emission beyond the LO Feynman diagram
in the PS and therefore regulates the  high-\pT\ emission against which the \ttbar\ system recoils.} was set to 1.5\,\mt~\cite{ATL-PHYS-PUB-2016-020}.
For the \ttbar process, the scales were set to $\muR^2 = \muF^2 = \mt^2 + \pTt^2$, where $\pTt$ is the \pT of the top quark, while for the \Wt and \sch processes these scales were set to \mt.
In the case of \Wt associated production, the diagram-removal scheme~\cite{Frixione:2008yi} was employed to handle the interference with \ttbar\ production~\cite{ATL-PHYS-PUB-2016-020}.
The \ttbar\ cross-section was calculated at NNLO in QCD including
resummation of next-to-next-to-leading logarithmic (NNLL) soft-gluon terms with \TOPpp[2.0]~\cite{Baernreuther:2012ws,Czakon:2011xx,Czakon:2012zr,Czakon:2012pz,Czakon:2013goa,Cacciari:2011hy}.
This cross-section is $832 \pm 35$~(PDF\,+\,\alphas)~${}^{+20}_{-29}$~(scale)~pb.
The cross-section uncertainties due to the PDF and \alphas\ were calculated using the PDF4LHC prescription with the \MSTW[nnlo] 68\%~CL, \CT[10nnlo]~\cite{Lai:2010vv} and \NNPDF[2.3nnlo] PDF sets, and were added in quadrature to the effect of the scale uncertainty.
The \Wt events are normalised to the predicted production
cross-section of $72\pm4$~pb calculated at NLO in QCD including
NNLL soft-gluon corrections~\cite{Kidonakis:2010ux}. The uncertainty in the cross-section
corresponds to the sum in quadrature of the uncertainty derived from
the \MSTW[nnlo] 90\%~CL PDF set and the effect of the scale uncertainties.
For the \sch\ process, the inclusive cross-section is corrected to the
theory prediction calculated at NLO in QCD with \HATHOR. This
cross-section is $10.3\pm0.4$~pb. The
uncertainties in the cross-section due to the PDF and \alphas\ were
calculated using the PDF4LHC prescription, similarly to the \tch.
 
Vector-boson production in association with jets (generally named $V$+\,jets, or $W$/\Zjets) was simulated with the multi-leg \SHERPA[2.2.1]~\cite{Bothmann:2019yzt} generator. The \NNPDF[3.0nnlo]
set~\cite{Ball:2014uwa} of PDFs as well as the dedicated set of tuned PS parameters developed by the \SHERPA authors for this version were used. The events were filtered according to their $b$-hadron and $c$-hadron content at the particle level. The ME+PS matching~\cite{Hoeche:2011fd} was employed for different jet
multiplicities, which were then merged into an inclusive sample using an improved CKKW matching
procedure~\cite{Catani:2001cc,Hoeche:2009rj} which is extended to NLO
accuracy using the MEPS@NLO prescription~\cite{Hoeche:2012yf}. These
particular simulations are accurate to NLO for up to two additional partons
and accurate to LO for up to four additional partons.
The virtual QCD corrections for MEs at NLO accuracy are
provided by the \OPENLOOPS library~\cite{Cascioli:2011va,Denner:2016kdg}. Fully leptonically and semileptonically decaying diboson ($VV$) samples were simulated with the \SHERPA[2.2.1] ME+PS generator. The ME+PS matching is the same as for the single-boson processes. These
particular simulations are accurate to NLO for up to one additional parton
and accurate to LO for up to three additional parton emissions using factorised
on-shell decays.
The inclusive cross-sections for $V$+\,jets production were calculated
to NNLO accuracy~\cite{ATL-PHYS-PUB-2017-006} with FEWZ program~\cite{Anastasiou:2003ds}. For \Wjets production, the overall theoretical uncertainty
is 34\%. This is the result of adding in quadrature the overall cross-section normalisation
uncertainty and 24\% per additional jet, according to the
Berends--Giele scaling~\cite{Berends:1990ax}.
 
The production of \ttZ, \ttW, \tZq, \tH, and \tWZ events was modelled by the \MGNLO[2.3.3] generator, which provides MEs at NLO in \alphas, with the \NNPDF[3.0nlo] PDF set. The production of \ttH\ events was modelled using the \POWHEGBOX generator at NLO with the \NNPDF[3.0nlo] PDF set.
 
All the signal and background processes involving one or more top quarks were simulated assuming a top-quark mass of \SI{172.5}{\GeV}, and the top quark was assumed to decay only into a $W$ boson and a $b$-quark.
In these samples, top quarks and $W$ and $Z$ bosons were decayed at LO using \MADSPIN~\cite{Frixione:2007zp,Artoisenet:2012st} to preserve all spin correlations.
Moreover, the PS, hadronisation and underlying-event (UE) modelling was simulated with \PYTHIA[8.230] or 8.212, using the A14 set of tuned parameters (A14 tune)~\cite{ATL-PHYS-PUB-2014-021} and the \NNPDF[2.3lo] PDF set.
The decays of bottom and charm hadrons were simulated using the \EVTGEN[1.6.0] or 1.2.0 program~\cite{Lange:2001uf}.
 
Alternative samples of simulated single-top-quark and \ttbar\ events were also produced, using either different generators or different values of parameters in \POWPY{8} (further details are given in \Section{sec:systematics}) to estimate the generator modelling uncertainties. For studies of the NLO matching method, \MGNLO[2.6.2] or 2.6.0 using either the \NNPDF[3.0nlo]{}~nf4 PDF set for the \tch\ process or the \NNPDF[3.0nlo] PDF set for the \ttbar, \tW\ and \sch\ processes, was used. In these cases, the ME generator was interfaced to \PYTHIA[8.230] or 8.212. To study the PS, the hadronisation and the UE modelling, the \POWHEGBOX[v2] generator interfaced to \HERWIG[7.04]~\cite{Bahr:2008pv,Bellm:2015jjp} using the H7UE set of tuned parameters~\cite{Bellm:2015jjp} and the \MMHT[lo] PDF set~\cite{Harland-Lang:2014zoa} was used. In the case of \Wt\ associated production, a simulated event sample using the diagram-subtraction scheme~\cite{Frixione:2008yi} is employed to estimate the uncertainty associated with the scheme used to handle the interference with \ttbar\ production.
 
Dijet events were simulated using \PYTHIA[8.186] with the A14 tune and the \NNPDF[2.3nlo] PDF set, and the decays of bottom and charm hadrons were simulated using the \EVTGEN[1.2.0] program. Here, $2 \rightarrow 2$ QCD processes were generated, including \qcd, $qg  \rightarrow q\gamma$, $q\bar{q}  \rightarrow g\gamma$, EW ($W$/$Z$) and \ttbar\ production processes. This simulated sample was filtered at generator level to enrich the event sample with jets that are likely to resemble electrons with detector signatures. Events were kept if particles in the event (excluding neutrinos and muons) deposit $>$\SI{17}{\GeV} of energy into a square area $\eta \times \phi = 0.1 \times 0.1$ of the electromagnetic calorimeter, mimicking the highly localised energy deposits characteristic of electrons.
 
All baseline simulated event samples were passed through the full simulation of the ATLAS detector~\cite{SOFT-2010-01} based on the \GEANT~\cite{Agostinelli:2002hh} framework. Simulated samples of fully polarised single top quarks and top antiquarks, used to estimate the impact of anomalous couplings or EFT on the differential measurements, and samples used to evaluate most of the systematic effects were processed with a fast simulation~\cite{SOFT-2010-01} which relies on a parameterisation of the calorimeter response~\cite{ATL-PHYS-PUB-2010-013}.
 
An extensive software suite~\cite{ATL-SOFT-PUB-2021-001} is used in the reconstruction and analysis of real and simulated data, in detector operations, and in the trigger and data acquisition systems of the experiment.


\section{Object definitions}
\label{sec:event-reconstruction}
Electron candidates are reconstructed from clusters of energy deposits in the electromagnetic calorimeter
that are matched to a track in the inner-detector tracking system. They are required to satisfy $\pt>\SI{7}{\GeV}$, $|\eta_{\text{cluster}}|<2.47$ and a `tight' likelihood-based identification requirement~\cite{EGAM-2018-01}. Electron candidates are excluded if their calorimeter
clusters lie within the transition region between the barrel and endcap sections of the electromagnetic calorimeter,
$1.37< |\eta_{\text{cluster}}|<1.52$. The track associated with the electron must pass the requirements $|z_0 \sin\theta| < 0.5$~mm
and $|d_0/\sigma{(d_0)} | < 5$, where $z_0$ is the longitudinal impact parameter with respect to the reconstructed
primary vertex, $d_0$ denotes the transverse impact parameter relative
to the beam-line axis and $\sigma{(d_0)}$ is the uncertainty in $d_0$.
 
Muon candidates are reconstructed from tracks measured in the muon
spectrometer matched to tracks measured in the inner-detector tracker in the pseudorapidity range of
$|\eta|<2.5$. They must satisfy $\pt>\SI{7}{\GeV}$ along with the `medium' identification requirements defined in
Ref.~\cite{MUON-2018-03}. These include requirements on the number of hits in the different inner-detector and muon
spectrometer subsystems
and on the significance of the charge-to-momentum ratio $q/p$. In addition, the track associated with the
muon candidate must have $|z_0 \sin\theta| < 0.5$~mm and $|d_0/\sigma{(d_0)} | < 3$.
 
Isolation criteria are applied to the selected electrons and muons.
For electrons, the scalar sum of the \pt of tracks within a variable-size cone around the electron,
excluding tracks originating from the electron itself, must be less than 6\% of the electron \pt.
The track isolation cone
size $\Delta R = \sqrt{(\Delta \eta)^2 + (\Delta \phi)^2}$ is given
by the smaller of $\Delta R = \SI{10}{\GeV}/\pT$ and $\Delta R = 0.2$. In addition, the sum of the transverse energy of the
calorimeter topo-clusters in a cone of $\Delta R = 0.2$ around the electron is required to be less than 6\% of the
electron $\pT$, excluding clusters originating from the electron itself. For muons, the scalar sum of the $\pT$
of tracks within a variable-size cone around the muon (excluding its own track) must be less than 6\% of
the muon $\pT$, with the track isolation cone size being given by
the smaller of $\Delta R = \SI{10}{\GeV}/\pT$ and
$\Delta R = 0.3$.
 
Jets are reconstructed from topological clusters of energy deposited in the calorimeter~\cite{PERF-2014-07}
using the \antikt\ algorithm with a radius
parameter of 0.4. They are calibrated through the application of a jet
energy scale derived from data and simulation~\cite{JETM-2018-05}.
These jets are required to have $\pT >\SI{30}{\GeV}$
and $|\eta|<4.5$. To suppress jets from additional $pp$ interactions
within a bunch crossing, the so-called jet-vertex tagger (JVT)~\cite{PERF-2014-03}  is
applied to jets with $\pt<\SI{120}{\GeV}$ and $|\eta|<2.5$.
Jets containing $b$-hadrons (named $b$-tagged jets) are identified (tagged)
by the MV2c10 $b$-tagging algorithm~\cite{ATL-PHYS-PUB-2017-013,FTAG-2018-01}.
The algorithm calculates a multivariate discriminant from information about the
impact parameters of associated charged-particle tracks, the properties of reconstructed
secondary vertices, and the topology of $b$- and $c$-hadron decays inside the jets.
The $b$-tagging efficiency is measured for jets
in the pseudorapidity range $|\eta|<2.5$ and with $\pT >
\SI{20}{\GeV}$ in simulated \ttbar\ events.
The chosen working point for this analysis corresponds to a
$b$-tagging efficiency of 60\%~\cite{PERF-2016-05}.
The corresponding mistagging rates for $c$-quark and light-flavour jets are
approximately 2.9\% and 0.065\%, respectively, as predicted in
simulated \ttbar\ events and calibrated in data~\cite{ATLAS-CONF-2018-001,ATLAS-CONF-2018-006}.
 
The missing transverse momentum in the event, whose magnitude is
denoted in the following by \MET, is defined as the negative vector
sum of the $\pt$ of the reconstructed and calibrated objects in the
event~\cite{PERF-2016-07,ATLAS-CONF-2018-023}. This sum also includes
a `soft term' consisting of the transverse momenta of inner-detector tracks that
are associated with the primary vertex but not with any other objects.
 
Objects can satisfy both the jet and lepton selection criteria and therefore a
procedure called `overlap removal' is applied to ensure that objects are matched to a unique hypothesis.
If any electron shares a track with a muon, the electron is removed since it is very likely to correspond to the reconstructed muon.
Similarly, if any jet is close to an electron (within $\Delta R < 0.2$), the closest jet is removed.
Electrons close to jets (within $\Delta R$ < 0.4) are also removed to reduce the impact of non-prompt leptons.
To reduce contributions from muons which stem from heavy-flavour decays inside a jet,
muons are removed if they are separated from the nearest jet by $\Delta R$ < 0.4.
Additionally, jets  with  fewer  than  three tracks  and  separated  from  a  muon  by $\Delta R$ < 0.4
are removed to reduce the number of fake jets from muons depositing a large fraction
of their energy in the calorimeters.

\section{Event selection in the signal and control regions}
\label{sec:event-selection}
 
Events are required to have at least one vertex reconstructed from at least two inner-detector tracks with
transverse momenta of $\pT>\SI{0.5}{\GeV}$. The primary vertex for
each event is defined as the vertex with the highest sum of \pTsq\ over all associated inner-detector tracks~\cite{ATL-PHYS-PUB-2015-026}.
The analysis considers only $W$-boson decay modes to an electron or a muon. Events in which the $W$ boson decays to a $\tau$-lepton are thus included if
the $\tau$-lepton subsequently decays to an electron or a muon.
The signal event candidates are selected by requiring a single
isolated lepton, significant \met\ and exactly two jets.  Each muon (electron)
is required to have $\pt>\SI{30}{\GeV}$ and $|\eta|<2.5$ ($|\eta|<2.47$, excluding the region $1.37< |\eta|<1.52$), and to satisfy
the identification and isolation criteria
discussed in \Section{sec:event-reconstruction}.
To remove background events from \ttbar, \Zjets and diboson production, the
event is vetoed if an additional `loose' lepton candidate with $\pt>\SI{10}{\GeV}$ is found
when applying less stringent lepton identification criteria and no isolation requirements~\cite{ATL-PHYS-PUB-2015-041,ATL-PHYS-PUB-2016-015,EGAM-2018-01}.
The value of \met is required to be larger than \SI{35}{\GeV}.
Jets in the endcap--forward transition region of the calorimeters, $2.75<|\eta|<3.5$, are
required to satisfy the more stringent requirement $\pt>\SI{35}{\GeV}$.  Exactly
one of the jets is required to be $b$-tagged and have $|\eta|<2.5$. The
other, non-$b$-tagged, jet is referred to as the `spectator jet'.
 
Two additional \qcd\ background rejection criteria are applied. The transverse mass of the lepton--$\MET$ system,
 
\begin{equation*}
\mtw = \sqrt{2 \pT \mbox{($\ell$)} \MET \left( 1 - \text{cos}~\Delta\phi ( \pT(\ell), \mbox{\MET} ) \right)} \,,
\end{equation*}
 
is required to be larger than \SI{60}{\GeV}, and $\Delta \phi(\pT(\ell), \MET)$ is the azimuthal angle between the lepton momentum and the \MET\ direction.
Further reduction of this background is achieved by imposing an additional requirement on events where the lepton and the leading jet in $\pT$ have opposite directions in the transverse plane,
 
\begin{equation}
\pT(\ell) > 50 \left( 1 - \frac{\pi - |\Delta \phi(\pT \mbox{($j_1$)}, \pT \mbox{($\ell$)})|}{\pi-1} \right)~\SI{}{\GeV} \,,
\label{eq:MultijetRejection2}
\end{equation}
where $\Delta \phi(\pT(j_1),\pT(\ell))$ is the azimuthal angle between the lepton \pT\ and the leading jet in \pT. This requirement provides significant rejection of background originating from \qcd\ events where two jets are produced back-to-back in the azimuthal plane and one of those is misreconstructed as a lepton.
This set of requirements defines the preselection region.
 
The on-shell $W$ boson originating from the decay of the top quark is
reconstructed from the momenta of the lepton and the neutrino by
imposing four-momentum conservation. Since the neutrino escapes
undetected, the $x$ and $y$ components of the reconstructed \MET are
assumed to correspond to the \pT\ of the neutrino. The unmeasured
longitudinal component of the neutrino momentum, $p^{\nu}_{z}$, is
computed by imposing a $W$-boson mass constraint on the
lepton--neutrino system. A quadratic expression is found for
$p^{\nu}_{z}$, and the solution closer to zero is taken. If the
solutions are complex, the reconstructed \MET\ is rescaled, preserving
its direction, in order to have a unique real solution for
$p^{\nu}_{z}$. The top-quark candidate is then reconstructed by
combining the four-momenta of the reconstructed $W$ boson and the
selected $b$-tagged jet.
Finally, the lepton momentum is boosted into the top-quark rest frame
and the angles $\theta_{\ell i}$ ($i = x', y', z'$) are derived by its
projection along the axes defined in \Section{sec:axes_definition}.
In doing so, the ambiguity in the direction of the incoming light
quark in the laboratory frame is resolved by accepting
the direction which is closer to that of the spectator quark, also in the
laboratory frame.
 
Further discrimination between \tch\ signal events and background events is achieved by applying additional criteria that optimise the signal-to-background ratio (S/B), and thereby determine the signal region. Thus, the following criteria are applied:
\begin{itemize}
\item The invariant mass of the lepton--$b$-tagged jet system, \mlb, is required to be lower than \cutMlb~\SI{}{\GeV}.
\item The mass of the reconstructed top quark, \mtop, is required to be within \cutMtopLower--\cutMtopUpper~\SI{}{\GeV}.
\item The mass of the spectator-jet--top-quark system, \mjt, is required to be larger than \cutMjt~\SI{}{\GeV}.

\item A trapezoidal requirement is imposed in order to reject background events from the two-dimensional correlation among the pseudorapidities of
the spectator jet, $\eta_j$, and the reconstructed top quark, $\eta_{\top}$. This requirement is:
\begin{eqnarray}
\eta_j &<& (4~\eta_{\top} + a)  \qquad \textsf{\&} \nonumber \\
\eta_j &>& (4~\eta_{\top} - a)  \qquad \texttt{\&} \nonumber \\
(\eta_j &>& (0.44~\eta_{\top} + b) \quad \textsf{OR} \nonumber \\
\eta_j &<& (0.44~\eta_{\top} - b)) \label{eq:trap}  \,,
\end{eqnarray}

where the intercept parameters $a$ and $b$ of the lines defining the trapezoid were optimised to be \trapcutA\ and \trapcutB, respectively.
 
\item The scalar sum of the \pT\ of all final-state objects, $\HT$, must be larger than \cutHT~\SI{}{\GeV}, since the $\HT$ distributions of the backgrounds peak at lower values than the \tch\ signature.
\end{itemize}

Additionally, two specific background-enriched control regions, orthogonal to the signal region, are defined in order to estimate the contributions of the most important background processes in the \tch\ signal region, coming from \ttbar\ and  \Wjets events, by computing scale factors for the overall normalisations.
These two specific background-enriched regions are:
 
\begin{itemize}
 
\item A control region enriched in \ttbar\ events is defined by applying all the preselection requirements, except for the requirement of exactly one $b$-tagged jet; instead, exactly two $b$-tagged jets are required.
 
\item A control region enriched in \Wjets events is defined by selecting events satisfying the preselection requirements
and at least one of the reversed requisites for \mlb, \mtop, \mjt\ or trapezoidal requirement, all from the
selection criteria.
This control region has a \Wjets flavour composition similar to that in the signal region (in terms of $W$+\,light-jets and $W$+\,heavy-jets contributions).
 
\end{itemize}
 
\Table{tab:summary_regions} summarises the selection criteria defining the preselection, the signal region and the two control regions used in this analysis. For the separate measurements of top-quark and top-antiquark events, the selected events in each region are further divided into two different regions according to the lepton charge.
 
\begin{table}[!htb]
\begin{center}
\begin{adjustbox}{max width=0.99\textwidth}
\begin{tabular}{p{3.6cm} | p{4.2cm} | p{3.6cm} | p{4.5cm}}
\hline
\hline
\multicolumn{4}{c}{Common event selection criteria} \\
\hline
\multicolumn{4}{c}{Exactly one electron or muon} \\
\multicolumn{4}{c}{Veto secondary low-\pT\ charged loose leptons} \\
\multicolumn{4}{c}{Exactly two jets} \\
\multicolumn{4}{c}{$\MET > \SI{35}{\GeV}$} \\
\multicolumn{4}{c}{$\mtw > \SI{60}{\GeV}$} \\
\multicolumn{4}{c}{$\pT(\ell) > 50 \left( 1 - \frac{\pi - |\Delta \phi(\pT \mbox{($j_1$)}, \pT \mbox{($\ell$)})|}{\pi-1} \right)~\SI{}{\GeV}$} \\
\hline
\hline
Preselection region & Signal region & \ttbar\ control region & \Wjets control region \\
\hline
Exactly one $b$-tagged jet & Exactly one $b$-tagged jet & Exactly two $b$-tagged jet & Exactly one $b$-tagged jet \\
& $\mlb < \cutMlb~\SI{}{\GeV}$ & & $\mlb > \cutMlb~\SI{}{\GeV}$ \\
& $\mjt > \cutMjt~\SI{}{\GeV}$ & & $\mjt < \cutMjt~\SI{}{\GeV}$ \\
& Trapezoidal requirement & & Veto trapezoidal requirement \\
& $\HT > \cutHT~\SI{}{\GeV}$ & & $\HT < \cutHT~\SI{}{\GeV}$ \\
\hline
\hline
\end{tabular}
\end{adjustbox}
\caption{Summary of the selection criteria defining the preselection, the signal region and the two control regions.}
\label{tab:summary_regions}
\end{center}
\end{table}


\section{Particle-level object definition and fiducial region selection}
\label{sec:particle-selection}
 
In order to reduce the dependency on phenomenological models which
describe colour reconnection, initial- and final-state radiation, and
fragmentation, the measured differential angular distributions are
unfolded to particle level.
 
\subsection{Particle-level objects}
 
The definitions of the particle-level objects
are the same as the ones detailed in Ref.~\cite{TOPQ-2015-05}. They are constructed from stable particles in
the MC event record with a lifetime larger than 30~ps, within the observable pseudorapidity range.
 
Particle-level leptons are defined as electrons, muons or neutrinos
that do not originate from hadron decays, either directly or via a
$\tau$-lepton decay. Thus, leptons from either a $W$- or $Z$-boson
decay are considered, including those emerging from a
subsequent $\tau$-lepton decay.
In \tch\ single-top-quark events, exactly one such electron or muon is present.
No isolation requirement is imposed on the
selected charged lepton and the calculation of its four-momentum includes photons
within a surrounding cone of size $\Delta R =0.1$.
The \MET\ is calculated from the vector sum of all the selected neutrinos.
 
Particle-level jets are reconstructed using the \antikt\ algorithm with
a radius parameter of 0.4. All stable particles are used to
reconstruct the jets, excluding electrons, muons, neutrinos, and
photons used in the definition of the selected charged leptons.
A particle-level jet is identified as a $b$-tagged jet if the jet is within
$|\eta| < 2.5$ and an associated $b$-hadron is found with a ghost-matching
technique~\cite{Cacciari:2008gn}; the hadron must have $\pt>\SI{5}{\GeV}$.
All particle-level charged leptons identified within a cone of size $\Delta R = 0.4$ around a selected particle-level jet are removed.

\subsection{Fiducial region definition}
 
The differential angular distributions are unfolded to particle level
in a fiducial region. This fiducial region is defined so as to be close to
the measured phase space, using the particle-level objects defined
in \Section{sec:event-reconstruction}. Exactly one particle-level
electron or muon with $\pt>\SI{30}{\GeV}$ and $|\eta|<2.5$ (excluding
the region $1.37< |\eta|<1.52$ for the case of electrons) is
required. There must be two particle-level jets with
$\pt>\SI{30}{\GeV}$; exactly one of these jets must be identified as a
$b$-tagged jet with $|\eta|<2.5$ while the other jet must satisfy
$|\eta|<4.5$. The particle-level \met\ is required to be larger than
\SI{35}{\GeV}. The two additional \qcd\ background rejection criteria,
i.e.\ \mtw\ larger than \SI{60}{\GeV} and the requirement in
\Equation{eq:MultijetRejection2}, and the remaining signal
region requirements are also applied
to the particle-level objects. The signal region criteria require the
reconstruction of the top quark. In this case, a top-quark proxy, called a pseudo top quark~\cite{TOPQ-2015-05}, is defined by using the particle-level
objects and following exactly the same method described in \Section{sec:event-selection}.
 
\Table{tab:fiducial_region} summarises the signal selection criteria,
applied to particle-level objects, for defining the fiducial region used in this analysis.
 
\begin{table}[!htb]
\begin{center}
\begin{adjustbox}{max width=0.99\textwidth}
\begin{tabular}{c}
\hline
\hline
Fiducial region \\
\hline
Exactly one electron or muon \\
Exactly two jets \\
Exactly one $b$-tagged jet \\
$\MET > \SI{35}{\GeV}$ \\
$\mtw > \SI{60}{\GeV}$ \\
$\pT(\ell) > 50 \left( 1 - \frac{\pi - |\Delta \phi(\pT \mbox{($j_1$)}, \pT \mbox{($\ell$)})|}{\pi-1} \right)~\SI{}{\GeV}$ \\
$\mlb < \cutMlb~\SI{}{\GeV}$ \\
$\mtop \in [\cutMtopLower, \cutMtopUpper]~\SI{}{\GeV}$ \\
$\mjt > \cutMjt~\SI{}{\GeV}$ \\
Trapezoidal requirement \\
$\HT > \cutHT~\SI{}{\GeV}$ \\
\hline
\hline
\end{tabular}
\end{adjustbox}
\caption{Summary of the signal selection criteria, applied to particle-level objects, for defining the fiducial region.}
\label{tab:fiducial_region}
\end{center}
\end{table}

\section{Background estimation}
\label{sec:background-estimation}
 
The largest background contributions to the fiducial region arise from
\ttbar\ and \Wjets production. The
former is difficult to distinguish from the \tch\ single-top-quark signal since
\ttbar\ events contain real top quarks in the final state. The
\Wjets production process contributes to the background if there is a
$b$-quark in the final state or if a $c$-jet or light-flavour jet is mistagged.
Other minor backgrounds originate from \Wt, \sch single-top-quark,
\Zjets, diboson, \ttZ, \ttW, \tZq, \tH, and \tWZ\ production. \QCD\
events produced via the strong interaction can also
contribute if, in addition to having two reconstructed jets, an extra
jet is misidentified as an isolated lepton, or if a non-prompt lepton
appears to be isolated.
 
For all processes except \qcd production, the
normalisations are initially estimated by using the simulated samples scaled
to the theoretical cross-section predictions for \pp\ collisions at
\cmenergy, discussed in \Section{sec:data_mc_samples}.
In the template fit and in the measurement of the angular differential
cross-sections, described in \Section{sec:template-fit}
and \Section{sec:unfolding-method}, respectively, the MC predicted yields for the two major background processes
(\ttbar and \Wjets) are normalised to the numbers of data events using the dedicated control regions
defined in \Section{sec:event-selection}.
The shape of the event distributions is taken from simulation.

The prediction for the normalisation and shape of any \qcd\ distribution is obtained by using the purely data-driven anti-muon model
for events containing a muon, and the mixed data--simulation
jet-electron model for events containing an
electron~\cite{ATLAS-CONF-2014-058, TOPQ-2011-02}.
In these methods,
event distribution templates are derived from data (for the anti-muon
method) or from dijet simulation (for the jet-electron method).
For the jet-lepton model, a dedicated selection is imposed on
the MC simulated dijet events, in order to enrich the sample with jets
that are likely to resemble a lepton in the detector.
The jet candidates are treated as a lepton henceforth. The anti-muon
model imposes a dedicated selection on data to enrich the sample with events that
contain fake muons.
The templates are then used in a binned maximum-likelihood fit to the \met\ distribution in the
electron channel, or to the $\mtw$ distribution in the muon
channel, in \qcd-enriched regions to extract the \qcd\ normalisation.
These two \qcd-enriched regions are
defined as the preselection region and the \ttbar\ control region
without applying the isolation requirement on the lepton \pT and
without applying the $\MET$ requirement for the electron channel and the
$\mtw$ requirement for the muon channel.
 

\section{Signal and background event yields}
\label{sec:event-yields}
 
\Table{tab:yields_prefit_combined} provides the predicted signal and background event yields for
the combined electron and muon channels in the preselection,
signal and control regions. The \qcd\ background is normalised as
explained in \Section{sec:background-estimation}.
Observed data yields are also shown.
In addition, the S/B ratio and the ratio of the observed number to the expected number of events (Data/Prediction) are shown for each region.
 
\begin{table}[!htb]
\centering
\begin{adjustbox}{max width=0.9\textwidth}

\begin{tabular}{l | D{,}{{}\pm{}}{-1} | D{,}{{}\pm{}}{-1} | D{,}{{}\pm{}}{-1} | D{,}{{}\pm{}}{-1}}
\hline
\hline
\multicolumn{1}{l|}{Process} & \multicolumn{1}{c|}{Preselection region} & \multicolumn{1}{c|}{Signal region} & \multicolumn{1}{c|}{$t\bar{t}$ control region} & \multicolumn{1}{c}{\Wjets control region} \\
\hline
$t$-channel & 219\,000,11\,000 & 70\,600,3500 & 13\,480,680 & 148\,200,7400 \\
$t\bar{t}$, $tW$, $s$-channel & 736\,000,39\,000 & 43\,200,2400 & 147\,800,8400 & 693\,000,37\,000 \\
\Wjets & 590\,000,200\,000 & 26\,200,8900 & 16\,100,5500 & 560\,000,190\,000 \\
\Zjets, diboson & 52\,900,5100 & 2120,350 & 2620,360 & 50\,800,4900 \\
Others & 494,38 & 30, 4 & 79, 6 & 464,36 \\
Multijet & 52\,000,10\,000 & 3500,640 & 5500,1800 & 48\,500,9400 \\
\hline
Total expected & 1\,650\,000,210\,000 & 145\,600,9900 & 186\,000,10\,000 & 1\,510\,000,200\,000 \\
\hline
Data &  \multicolumn{1}{c|}{1\,750\,918} &  \multicolumn{1}{c|}{154\,361} &  \multicolumn{1}{c|}{188\,326} &  \multicolumn{1}{c}{1\,596\,557} \\
\hline
\hline
S/B &  0.15,0.02 &  0.94,0.13 &  0.08,0.01 &  0.11,0.02 \\
\hline
Data/Prediction &  1.06,0.13 & 1.06,0.07 & 1.02,0.06 & 1.06,0.14 \\
\hline
\hline
\end{tabular}

\end{adjustbox}
\caption{Pre-fit event yields in the preselection and signal regions
and in the \ttbar\ and \Wjets control regions for the combined electron and muon channels. The predictions are derived from
simulated event samples normalised to the theoretical cross-sections. For \qcd\ production the normalisation is estimated using a data-driven likelihood fit.
The label `Others' represents \ttZ, \ttW, \tZq, \tH, and \tWZ\ production.
The data-driven scale factors obtained for the top-quark and
\Wjets background processes are not considered when
computing these event yields.
The uncertainties shown account for
systematic effects and the uncertainty due to limited MC
sample size.
The expected S/B ratio and the ratio of the observed number to the expected number of events are also given.}
\label{tab:yields_prefit_combined}
\end{table}

\section{Sources of systematic uncertainty}
\label{sec:systematics}

Various sources of systematic uncertainty affect the signal and
background rates and the shape of the kinematic and angular
distributions.

In the following, the procedures used to evaluate the systematic uncertainties are described.
The systematic uncertainties are grouped into two main categories: experimental uncertainties
and theoretical modelling uncertainties. The effect
due to the limited size of the simulated event samples is also taken
into account when evaluating the total uncertainty.

\textbf{Experimental uncertainties:}
The uncertainty in the combined \datafirstyear--\datalastyear\
integrated luminosity is \lumiRelUnc\%~\cite{ATLAS-CONF-2019-021},
obtained using the LUCID-2 detector~\cite{LUCID2} for the primary luminosity measurements.
To account for the difference between the pile-up distributions in data
and MC simulations, an uncertainty related to the scale factors used
to adjust the MC pile-up to the data pile-up profile is applied.

For electrons and muons, the reconstruction,
identification, isolation, and trigger performance
can differ slightly between data and MC simulation. Scale factors are applied to simulated events to correct for these differences.
These scale factors, estimated using the tag-and-probe
method~\cite{EGAM-2018-01,MUON-2018-03}, as well as the lepton momentum
scale and resolution, are assessed using leptonic decays of $Z$ bosons and \Jpsi mesons.
Corrections to the lepton momentum scale and resolution are
applied to data.
The associated systematic uncertainties are then propagated to the
distributions used in this analysis.

To determine the jet energy scale (JES) uncertainty, information from test-beam data, LHC collision data, and
simulation was used, as described in Ref.~\cite{JETM-2018-05}.
The JES uncertainty is decomposed into a set of 29 uncorrelated
components, with contributions from pile-up, jet
flavour composition, single-particle response, and effects of jets not
contained within the calorimeter.
The jet energy resolution (JER) is measured separately for
data and MC simulation, using in situ techniques~\cite{JETM-2018-05}. The measured relative JER ranges from 22\% at \SI{30}{\GeV} to 6\% at \SI{300}{\GeV}.
Its uncertainty is represented by eight components
accounting for jet-$\pT$ and $\eta$-dependent differences between
simulation and data.
In the measurement of the three components of the polarisation vector,
an additional uncertainty is considered in order to take into account
the kinematic and sample dependence of the JER uncertainty model. Further details are given in
\Section{sec:template-fit}.
The systematic uncertainty associated with the
JVT is obtained by increasing and decreasing the scale factor used to correct the JVT
efficiency in simulation within its
uncertainties~\cite{PERF-2014-03}.

The $b$-tagging efficiency and mistagging rates are measured in data
using the methods as described in Refs.~\cite{FTAG-2018-01, ATLAS-CONF-2018-001,ATLAS-CONF-2018-006}, with the
systematic uncertainties due to the $b$-tagging efficiency and the
mistagging rates calculated separately. The impact of the
uncertainties on the $b$-tagging calibration is evaluated separately for
$b$-jets, $c$-jets and light-flavour jets.

The systematic uncertainties related to the modelling of the \MET\ in
the simulation are estimated by propagating the uncertainties in the
energy and momentum scales of electrons, muons and jets, as well as
the uncertainties in the resolution and scale of the soft term~\cite{PERF-2016-07}.

\textbf{Theoretical modelling uncertainties:}
Systematic uncertainties associated with the signal and background MC
modelling are estimated by comparing event samples from different
generators and by varying parameter values in the event generation.

To assess the uncertainty due to the choice of matching scheme in
the \tch\ single-top-quark signal and single-top-quark background ME generation, samples from the nominal MC generator
\POWHEGBOX\ are compared with those from \MGNLO, both interfaced to \PYTHIA[8].
 
The uncertainty in the PS is estimated for all top-quark processes by comparing samples from \POWHEGBOX interfaced to \PYTHIA[8] with
samples from \POWHEGBOX interfaced to \HERWIG[7].
 
The uncertainty due to missing higher-order QCD corrections in the ME
computation is estimated for all top-quark processes by independently
varying the renormalisation
and factorisation scales by factors of 0.5 and 2.0 from their
central value. Additionally, uncertainties due to initial-state radiation from the PS are assessed by
varying the corresponding parameter of the A14 parton
shower tune \texttt{Var3c}~\cite{ATL-PHYS-PUB-2014-021}.
For the \ttbar\ process and just in the case of the upward variation, the \hdamp\ parameter is also changed and set to 3\,\mt~\cite{ATL-PHYS-PUB-2020-023}.
The uncertainties due to final-state radiation from the PS are assessed by
varying the corresponding parameter of the A14 parton
shower tune \texttt{Var2}~\cite{ATL-PHYS-PUB-2014-021} and by varying
the renormalisation scale for QCD emission by factors of 0.5 and 2.0.
 
An additional uncertainty arising from the method used to handle the
interference between \tW\ and \ttbar\ production is determined by
comparing the \tW\ simulated sample that uses the diagram-subtraction
method with the nominal one based on the diagram-removal technique.

In the template fit, described in \Section{sec:template-fit}, an additional uncertainty due to \tch\
modelling arises from the use of a LO generator (i.e.\ \PROTOS) for the
construction of templates in the signal region as detailed
in \Section{sec:template-fit}, a choice which was motivated by the possibility
of varying the polarisation in a straightforward way. This is
determined by comparing the LO \PROTOS prediction with the NLO \POWHEGBOX\ prediction.
 
PDF uncertainties are evaluated using the PDF4LHC15 uncertainty set which
consists of 30 eigenvector variations from multiple NLO PDF sets~\cite{Butterworth:2015oua}.

The event yields associated with the simulated signal and background
processes are estimated using the selection acceptances and the
theoretically predicted cross-sections as reported
in \Section{sec:data_mc_samples}. The uncertainties in these
cross-sections are taken into account. However, for those processes
whose normalisation is extracted from the data-driven fits,
described in Sections~\ref{sec:template-fit}
and~\ref{sec:unfolding-method}, the effect is negligible.
The \qcd\ background is normalised
through a data-driven analysis based on the techniques described in \Section{sec:background-estimation}. A relative systematic
uncertainty of 20\% (40\%) is assigned to this data-driven overall
normalisation in the signal and \Wjets-dominated (\ttbar-dominated)
regions. It is estimated from fits, as described in \Section{sec:background-estimation}, using alternative kinematic variables
(\HT\ and $\Delta \phi(\pT(j_1),\pT(\ell))$ in the electron channel and $\Delta \phi(\pT(\ell), \MET)$ in the muon channel) and including modelling
uncertainties for the main background processes.

To evaluate the impact of the systematic uncertainty
in the shape of the \Wjets simulated templates, an upward and downward variation of
30\% is applied independently to the $W$+\,$b$-jets and $W$+\,$c$-jets
samples~\cite{HIGG-2016-29}. 
No variation is considered for
$W$+\,light-flavour jets production since its contribution is negligible.
To evaluate the systematic uncertainty in the shape of the \qcd\
templates, additional MC simulation and data-driven samples were produced, varying the fraction of the
jet's energy that is deposited in the EM calorimeter (jet-electron
model) and the ratio of the sum of the transverse momenta of tracks within a cone of maximum size $\Delta R = 0.4$
to the \pT\ of the muon (anti-muon model). These
alternative \qcd\ templates are normalised to the nominal yields and
compared with the nominal \qcd\ templates.

\textbf{Statistical uncertainties:}
Statistical fluctuations in the MC simulated event samples contribute to the overall systematic
uncertainty. These uncertainties arise from the finite number of
simulated background and signal events.
This uncertainty is discussed further in Sections~\ref{sec:template-fit} and~\ref{sec:unfolding-method}.

\section{Measurement of top-quark and top-antiquark polarisation vectors}
\label{sec:template-fit}
 
\begin{figure}[!htpb]
\begin{center}
\includegraphics[height=.30\textheight]{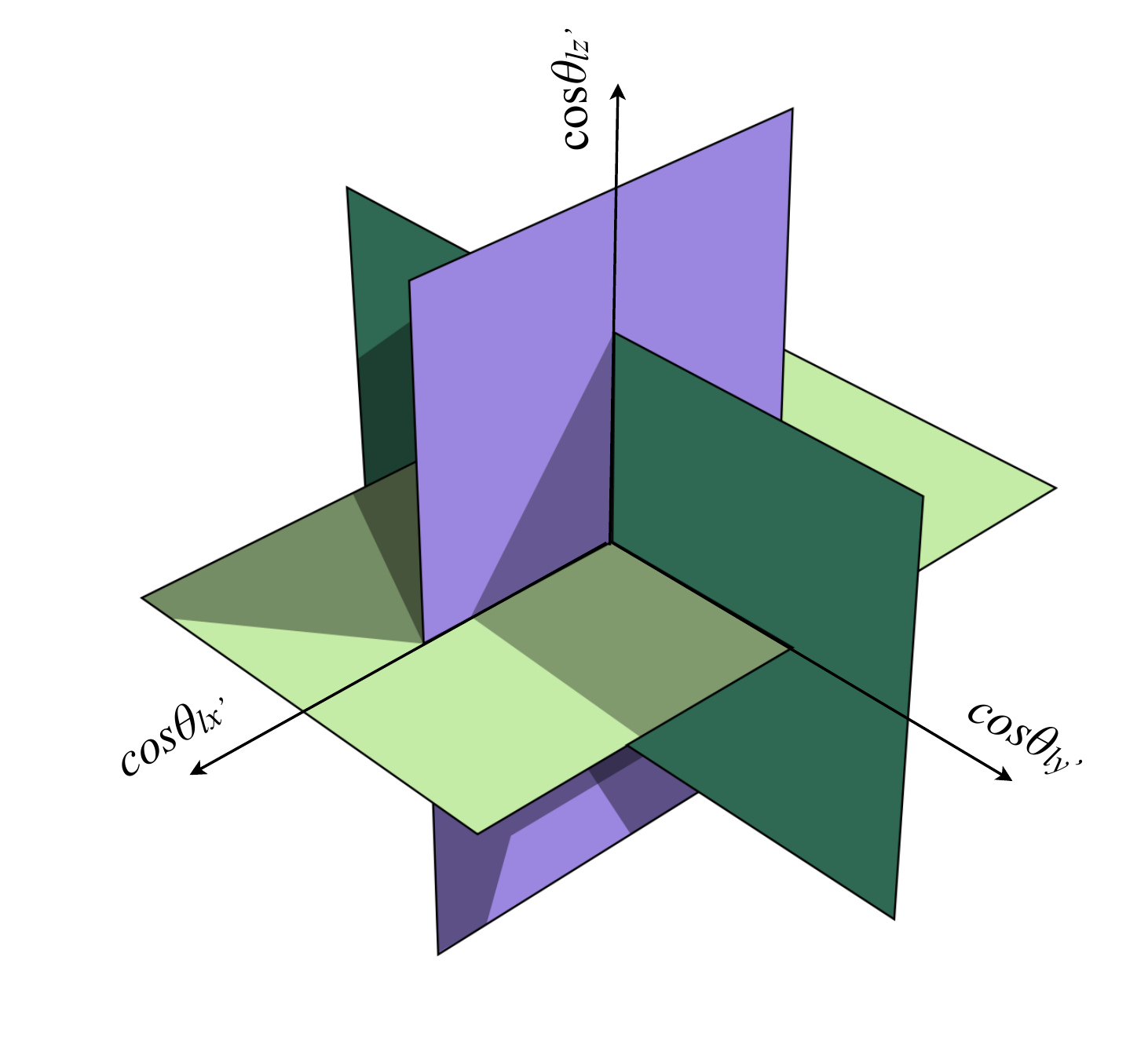}
\end{center}
\caption{
Representation of the octant variable $Q$ constructed by
slicing the tridimensional phase space into eight octants, according
to the signs of the three variables $\costhetax$, $\costhetay$,
$\costhetaz$.\label{fig:Octants}}
\end{figure}

The method used to measure the polarisation vector of an ensemble of
top quarks or antiquarks that pass the event selection is discussed in this section. The unit vector of the reconstructed
lepton momentum in the top-(anti)quark reference frame is first determined in the coordinate
system described in \Section{sec:axes_definition}.  The variable $Q$
(also called the \textit{octant variable}) is constructed by
slicing the tridimensional phase space into eight octants, in terms of the signs of three variables
$\costhetax$, $\costhetay$, $\costhetaz$ of this unit vector as
illustrated in \Figure{fig:Octants}. An analysis of the
coursely-binned fully differential top-quark decay distribution
described in Ref.~\cite{Aguilar-Saavedra:2017wpl} is afforded by the variable
$Q$. Mathematically, it is assigned a value zero through seven according to the equation
$Q=4\cdot\Theta(\costhetaz)+2\cdot\Theta(\costhetax)+\Theta(\costhetay)$
where $\Theta(\xi)$ is the Heaviside, or unit, step function of the variable $\xi$.
The signal region is further divided by lepton
charge for a total of 16 bins indexed by $Q_+=0,...,7$ for bins
of positive lepton charge and $Q_-=0,...,7$ for bins of negative lepton charge.
Four bins of control data consisting of events with positive
and negative leptons in the \Wjets and the \ttbar\ control
regions, as described in~\Section{sec:event-selection}, are
also considered.
These bins, related to the background control region, are indexed by $\mathcal{R}_{\pm}\in\left\{\Wjets,t\bar{t}\right\}$ where the sign corresponds to the sign of the charged lepton's charge.
This allows better control of these important backgrounds in the
\tch\ signal region. The event count in these 20 bins constitutes
the twenty-bin input data distribution to a binned profile-likelihood fit which is designed to extract the polarisation vector of the top-quark and top-antiquark event sample.
 
A parameterised fitting function is constructed to describe the distribution in the variable $Q_\pm$ of the input data.
Adjustable parameters of the function include $\vec{P}=\{\Polx,\Poly,\Polz\}$, separately for top
quarks and top antiquarks (i.e.\ six parameters in total).  Nuisance
parameters (NPs) include the overall normalisation factors $\beta_k$ for the production of the process $k\in$ (\tch, \ttbar, \Wjets), as well
as other NPs,  collectively denoted by $\vec{\theta}$, corresponding to the systematic uncertainties described
in \Section{sec:systematics}. The fitting function in the variable $Q_\pm$ is
determined by the projection of the joint probability density of the four-dimensional fully differential angular decay distribution for the
top-(anti)quark decays~\cite{Aguilar-Saavedra:2017wpl} onto the octant variable $Q_\pm$.
It is obtained from templates derived from histograms of quantities in events produced using a LO generator (i.e.\ \PROTOS).
These templates $\mathcal{T}_i\left(Q_\pm\right)$, with $i \in \left\{z'_+, z'_-,x',y'\right \}$,
are produced using samples of fully simulated events in which the polarisation of the top (anti)quarks
is manipulated to lie fully along the positive or negative $x'$, $y'$, or $z'$ directions.
The events are subjected to the same reconstruction procedures and event selection as the data.
 
The fitting function $\mu\left(Q_\pm \right)$ for the data is built from these templates plus an additional contribution from background events, also
estimated from reconstructed quantities. It describes the expected number of signal events falling within a bin $Q_\pm$ and is written as
\small
\begin{equation*}
\mu\left(Q_{\pm};\vec{P},\vec{\beta},\vec{\theta}\right)  = \beta_{t\text{-channel}}\cdot \Big\{ \frac{1+\Polz}{2}\mathcal{T}_{z'_+}\!\left(Q_{\pm}\right) + \frac{1-\Polz}{2}\mathcal{T}_{z'_-}\!\left(Q_{\pm}\right)
+ \frac{\Polx}{2}\mathcal{T}_{x'}\!\left(Q_{\pm}\right)
+ \frac{\Poly}{2}\mathcal{T}_{y'}\!\left(Q_{\pm}\right)  \Big\}
+ \mathcal{T}_{\text{bkg}}\!\left(Q_{\pm};\beta_{\Wjets},\beta_{t\bar{t}}\right).
\end{equation*}
\normalsize
Here,
$\mathcal{T}_{\text{bkg}}\left(Q_{\pm};\beta_{\Wjets},\beta_{t\bar{t}}\right)$ is the template for the background, consisting of a sum over all sources of background. Likewise, the number of events in each of the four bins of control data is described by the function $\nu(\mathcal{R}_{\pm};\vec{\beta},\vec{\theta})$, also derived either from simulation or data-driven methods.
 
The fitting function is used to construct a likelihood function
\begin{eqnarray*}
\mathcal{L\left( \vec{P},\vec{\beta},\vec{\theta}  \right)}
&=& \prod_{Q_+ = 0}^{7} \mathcal{P}\left(N_{Q_+};\mu\left(Q_+;\vec{P},\vec{\beta},\vec{\theta}\right)\right) \prod_{Q_- = 0}^{7} \mathcal{P}\left(N_{Q_-};\mu\left(Q_-;\vec{P},\vec{\beta},\vec{\theta}\right)\right) \nonumber \\
&\times & \prod_{\mathcal{R}_{+}\in\left\{\Wjets,t\bar{t}\right\}}\mathcal{P}\left(N_{\mathcal{R}_+};\nu\left(\mathcal{R}_+;\vec{\beta},\vec{\theta}\right)\right)\prod_{\mathcal{R}_{-}\in\left\{\Wjets,t\bar{t}\right\}}\mathcal{P}\left(N_{\mathcal{R}_-};\nu\left(\mathcal{R}_-;\vec{\beta},\vec{\theta}\right)\right) \nonumber \\
&\times& \prod_l \mathcal{G} (\theta_l;0,1) \,,
\end{eqnarray*}
in which the first line refers to the octants, the second line refers to the control regions,  and the third line refers to Gaussian constraints
on the NPs $\vec{\theta}$, where $\mathcal{P}\left(N;\lambda \right)$
indicates a Poisson distribution for $N$ events given expectation $\lambda$, and where $\mathcal{G}\left(m;m_0,\sigma\right)$ indicates a normal distribution in the variable $m$ with mean $m_0$ and standard deviation $\sigma$.
The full likelihood including all NPs is maximised
to extract the six components of $\vec{P}^{t}=
\{\Polx^{t},\Poly^{t},\Polz^{t}\}$ and $\vec{P}^{\bar{t}}= \{\Polx
^{\bar{t}},\Poly ^{\bar{t}},\Polz ^{\bar{t}}\}$ for top quarks and
antiquarks, respectively.
Nuisance parameters accounting for systematic uncertainties are not
considered in the fit if they have an impact on either normalisation or shape which is below 0.5\%.

The angular distributions studied in this paper are sensitive to the
JER. The corrections and uncertainties in jet energy are \pT\ and $\eta$
dependent and were determined with in situ techniques using dijet
events. Poorer JER was observed in simulated single-top-quark \tch\
events than in simulated dijet events.
This increase in JER value, which may be attributed to different event kinematics,
is also observed in the template fit.

\begin{figure}[!htpb]
\begin{center}
\subcaptionbox{}{\includegraphics[height=0.35\textheight]{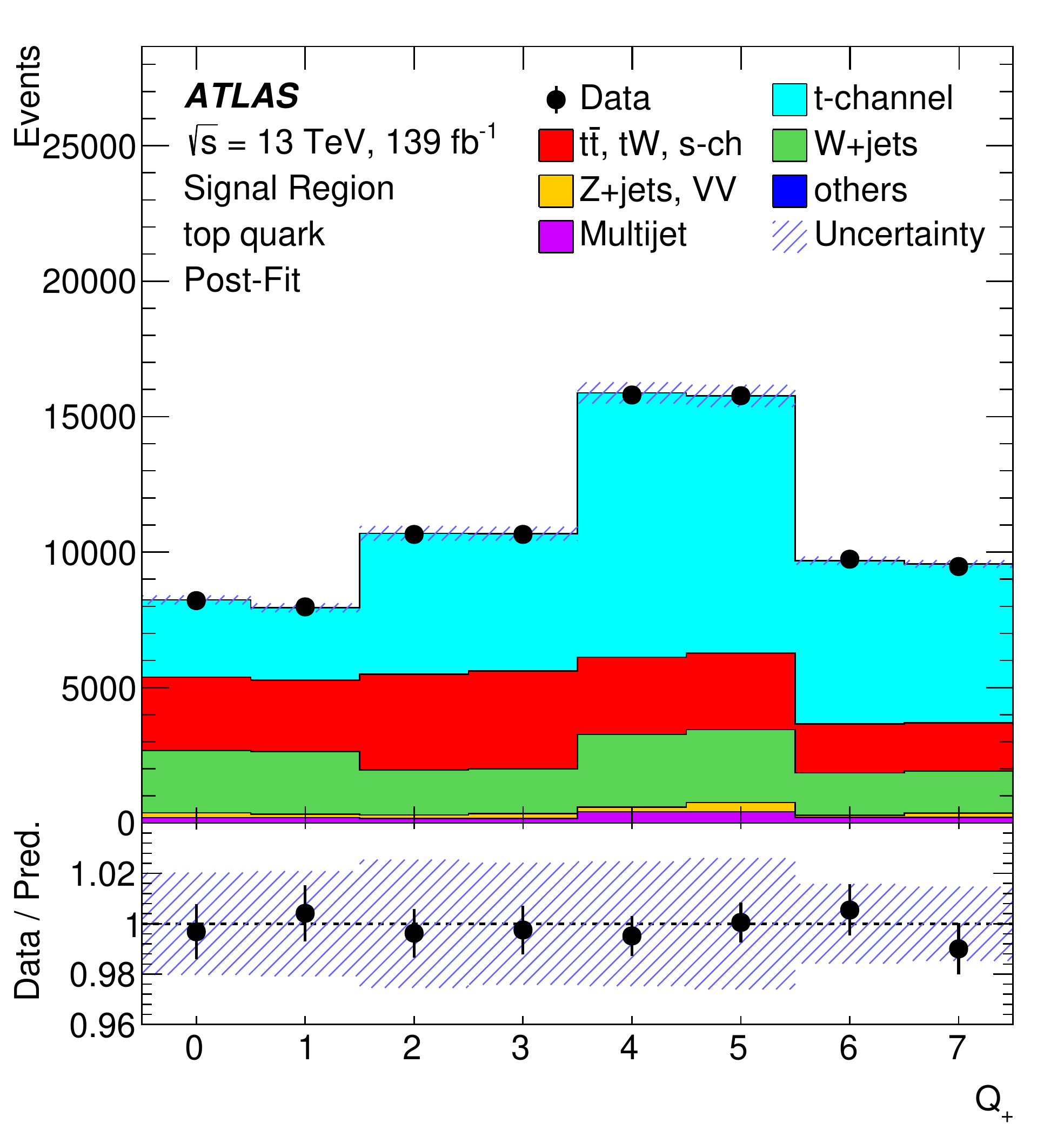}}
\subcaptionbox{}{\includegraphics[height=0.35\textheight]{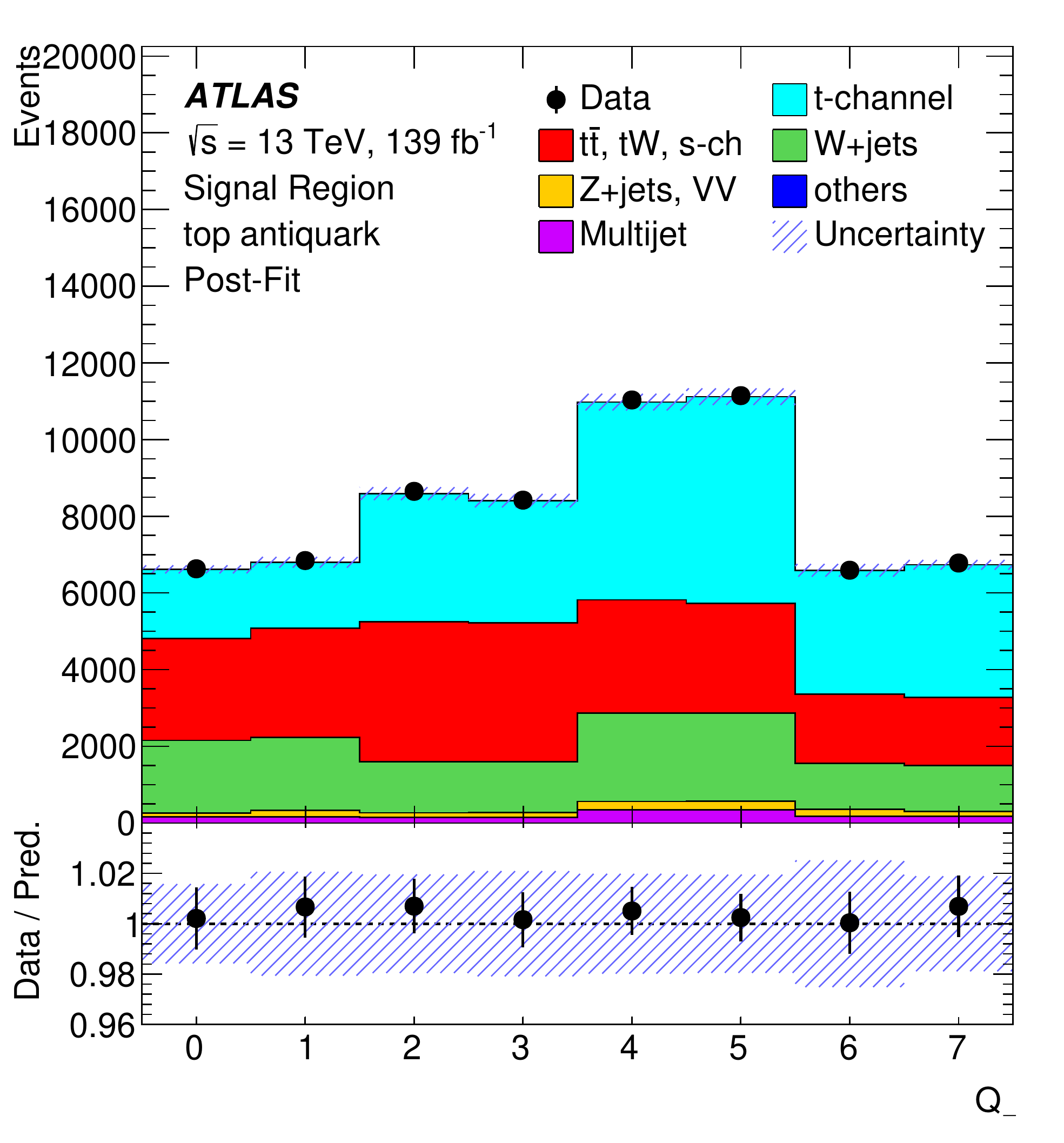}}
\end{center}
\caption{
Observed data and fitted distributions of
the octant variable (a) ${Q_+}$ in the top-quark and (b) ${Q_-}$
in the top-antiquark signal regions.
The $Q$ variable is assigned an integer value zero through seven according to
$Q=4\cdot\Theta(\costhetaz)+2\cdot\Theta(\costhetax)+\Theta(\costhetay)$
where $\Theta(\xi)$ is the Heaviside step function of the
variable $\xi$.
The label `others' represents \ttZ, \ttW, \tZq, \tH, and \tWZ\ production.
The uncertainty bands include both the statistical and systematic
uncertainties.
The lower panels show the ratio of data to prediction in each bin. \label{fig:SR_fit_data} }
\end{figure}

In order to account for the difference, two JER uncertainty models are
compared. The nominal fit model is used to obtain the central values
for the polarisation. In this model, the JER is allowed to vary
independently for each bin of the octant variable and for each control
region. With this approach, the role of the single-top-quark \tch\
events in constraining the JER is reduced. In the second model, \tch\
events are allowed to have a larger impact on the JER. This is
achieved by allowing each uncertainty affecting the JER to vary in a
correlated way across all bins, leading to the larger JER value
mentioned above. The polarisation is
measured again in this second fit. The difference between the polarisations obtained with the
two fit models is added in quadrature to the uncertainty obtained from
the nominal model in order to get the overall uncertainty for the
measurement.
This additional uncertainty is more pronounced
for \Polx,  while \Polz and \Poly  are much less affected.
\Figure{fig:SR_fit_data} shows the observed and fitted numbers of
events per octant $Q$ for top quarks and top antiquarks, separately, after the
nominal fit to data.
In this figure, neighboring bins refer to pairs of quadrants lying on opposite sides of the production plane, i.e. differing by the sign of $\costhetay$.
The first four bins indicate leptons emitted opposite to the spectator-quark
direction in the top-quark reference frame, while the last four bins refer
to leptons emitted along the spectator-quark direction, i.e. differing by the sign of $\costhetaz$.

The extracted polarisations together with the normalisation factors of
the \tch, \ttbar\ and \Wjets processes are shown in
\Table{tab:data_polarisations}.
The column labelled `extracted value' lists the result for the three
normalisations and the six polarisation components, and the total
uncertainty for each. In the column labelled `(stat.)' the statistical
uncertainty is presented separately.
These may be compared with parton-level predictions calculated at NNLO, based on
Ref.~\cite{Berger:2017zof}, where stable
single top quarks produced in the \tch\ from $pp$ collisions at
\cmenergyThirteen\ are considered.
These predictions are $\Polx^t= -0.024 \pm 0.001 ~\text{(scale)}{}^{+0.004}_{-0.007} ~\text{(PDF\,+\,\alphas)}$
and $\Polz^t=0.965 \pm{0.003} ~\text{(scale)}{}^{+0.003}_{-0.004} ~\text{(PDF\,+\,\alphas)}$ for top quarks
and $\Polx^{\bar t}= -0.073 \pm 0.008 ~\text{(scale)}{}^{+0.008}_{-0.004} ~\text{(PDF\,+\,\alphas)}$
and $\Polz^{\bar t}=-0.957 {}^{-0.003}_{+0.012} ~\text{(scale)}{}^{+0.004}_{-0.002} ~\text{(PDF\,+\,\alphas)}$ for top antiquarks. Because of
CP symmetry in top-quark production, $\Poly^t$ and $\Poly^{\bar t}$ are expected to be zero~\cite{Aguilar-Saavedra:2014eqa}; equivalently,
the polarisation vector is expected to lie in the plane of production, as described in \Section{sec:intro}.
The only non-zero SM contributions to $\Poly^t$ and $\Poly^{\bar{t}}$ come from absorptive parts of EW NLO diagrams, which are much smaller than the expected uncertainty of this measurement. These theoretical predictions and their uncertainties are calculated as detailed in \Section{sec:intro}.
 
A summary of the impact of the systematic uncertainties on the value of each polarisation parameter is shown in \Table{tab:systs_group_data}.
The larger cross-section for top-quark versus top-antiquark production
leads to smaller uncertainties for the measured polarisation for top quarks.
Systematic effects tend to have symmetric impacts for either side of the production plane, leading to reduced impact on $\Poly$ versus the other directions.
The uncertainties with the largest impact are those due to the JES and JER. This is because the polarisation depends on kinematic angles
determined in the top-quark reference frame, and since jet energy and
\MET are used to reconstruct that frame, uncertainties in the
measurement of the jet energy
are expected to contribute significantly to the total systematic
uncertainty.
The finite number of simulated events as well as the \ttbar
modelling are also important sources of uncertainty.
The normalisation factor extracted for the \tch\ signal contribution is
compatible with results obtained in Ref.~\cite{TOPQ-2015-15}.
 
\begin{table}[!htbp]
\centering
\begin{adjustbox}{max width=1.0\textwidth}

\begin{tabular}{l c c}
\hline
\hline
\noalign{\vskip 0.8mm}
Parameter & Extracted value & (stat.) \\
\noalign{\vskip 0.3mm}
\hline
\noalign{\vskip 1.0mm}
$t$-channel norm. & $+1.045\pm0.022$ & (${}\pm0.006$) \\
\noalign{\vskip 1.0mm}
\Wjets norm. & $+1.148\pm0.027$ & (${}\pm0.005$) \\
\noalign{\vskip 1.0mm}
$t\bar{t}$ norm. & $+1.005\pm0.016$ & (${}\pm0.004$) \\
\noalign{\vskip 1.0mm}
\hline
\noalign{\vskip 1.0mm}
${P}_{x'}^{t}$ & $+0.01~~\pm 0.18~~$ & (${}\pm0.02~~$) \\
\noalign{\vskip 1.0mm}
${P}_{x'}^{\bar{t}}$ & $-0.02~~\pm 0.20~~$ & (${}\pm0.03~~$) \\
\noalign{\vskip 1.0mm}
$P_{y'}^{t}$ & $-0.029\pm0.027$ & (${}\pm0.011$) \\
\noalign{\vskip 1.0mm}
${P}_{y'}^{\bar{t}}$ & $-0.007\pm0.051$ & (${}\pm0.017$) \\
\noalign{\vskip 1.0mm}
${P}_{z'}^{t}$ & $+0.91~~\pm 0.10~~$ & (${}\pm0.02~~$) \\
\noalign{\vskip 1.0mm}
${P}_{z'}^{\bar{t}}$ & $-0.79~~\pm0.16~~$ & (${}\pm0.03~~$) \\
\noalign{\vskip 1.0mm}
\hline
\hline
\end{tabular}

\end{adjustbox}
\caption{\label{tab:data_polarisations} Normalisation factors of the \tch,
\Wjets and \ttbar\ processes together with the polarisation values as
extracted from data, including total and statistical-only uncertainties in the fit.
}
\end{table}
 
\begin{table}[!htbp]
\centering
\begin{adjustbox}{max width=1.0\textwidth}

\begin{tabular}{l c c c c c c}
\hline
\hline
\noalign{\vskip 0.8mm}
Uncertainty source & $\Delta{P_{x'}^{t}}$ & $\Delta{P_{x'}^{\bar{t}}}$ & $\Delta{P_{y'}^{t}}$ & $\Delta{P_{y'}^{\bar{t}}}$ & $\Delta{P_{z'}^{t}}$ & $\Delta{P_{z'}^{\bar{t}}}$ \\
\noalign{\vskip 0.3mm}
\hline
\textbf{Modelling} &&&&&& \\
\noalign{\vskip 0.4mm}
Modelling ($t$-channel) & $\pm0.037$ & $\pm0.051$ & $\pm0.010$ & $\pm0.015$ & $\pm0.061$ & $\pm0.061$ \\
\noalign{\vskip 0.4mm}
\noalign{\vskip 0.4mm}
Modelling ($t\bar{t}$) & $\pm0.016$ & $\pm0.021$ & $\pm0.004$ & $\pm0.016$ & $\pm0.003$ & $\pm0.016$ \\
\noalign{\vskip 0.4mm}
\noalign{\vskip 0.4mm}
Modelling (other) & $\pm0.013$ & $\pm0.031$ & $\pm0.003$ & $\pm0.006$ & $\pm0.026$ & $\pm0.043$ \\
\noalign{\vskip 0.4mm}
\hline
\textbf{Experimental} &&&&&& \\
\noalign{\vskip 0.4mm}
Jet energy scale & $\pm0.045$ & $\pm0.048$ & $\pm0.005$ & $\pm0.007$ & $\pm0.033$ & $\pm0.025$ \\
\noalign{\vskip 0.4mm}
\noalign{\vskip 0.4mm}
Jet energy resolution & $\pm0.166$ & $\pm0.185$ & $\pm0.021$ & $\pm0.040$ & $\pm0.070$ & $\pm0.130$ \\
\noalign{\vskip 0.4mm}
\noalign{\vskip 0.4mm}
Jet flavour tagging & $\pm0.004$ & $\pm0.002$ & $<$$0.001$ & $\pm0.001$ & $\pm0.007$ & $\pm0.009$ \\
\noalign{\vskip 0.4mm}
\noalign{\vskip 0.4mm}
Other experimental uncertainties & $\pm0.015$ & $\pm0.029$ & $\pm0.002$ & $\pm0.007$ & $\pm0.014$ & $\pm0.026$ \\
\noalign{\vskip 0.4mm}
\noalign{\vskip 0.4mm}
Multijet estimation & $\pm0.008$ & $\pm0.021$ & $<$$0.001$ & $\pm0.001$ & $\pm0.008$ & $\pm0.013$ \\
\noalign{\vskip 0.4mm}
\noalign{\vskip 0.4mm}
Luminosity & $\pm0.001$ & $\pm0.001$ & $<$$0.001$ & $<$$0.001$ & $<$$0.001$ & $<$$0.001$ \\
\noalign{\vskip 0.4mm}
\noalign{\vskip 0.4mm}
Simulation statistics & $\pm0.020$ & $\pm0.024$ & $\pm0.008$ & $\pm0.015$ & $\pm0.017$ & $\pm0.031$ \\
\noalign{\vskip 0.4mm}
\hline
\noalign{\vskip 0.4mm}
\textbf{Total systematic uncertainty} & $\pm0.174$ & $\pm0.199$ & $\pm0.025$ & $\pm0.048$ & $\pm0.096$ & $\pm0.153$ \\
\textbf{Total statistical uncertainty} & $\pm0.017$ & $\pm0.025$ & $\pm0.011$ & $\pm0.017$ & $\pm0.022$ & $\pm0.034$ \\
\noalign{\vskip 0.4mm}
\hline
\hline
\end{tabular}

\end{adjustbox}
\caption{\label{tab:systs_group_data}
Systematic and statistical uncertainties in the measurement of the polarisation vector $\vec{P}$ for top quarks and top antiquarks.
The impact of each group of uncertainties is obtained by performing a fit where the NPs
in the group are fixed to their best-fit values,
subtracting the square of the resulting uncertainty in the parameter of
interest (i.e.\ for each polarisation component)
from the squared uncertainty from the nominal fit, and then taking the square root.
An additional uncertainty from the JER is included, consisting of
the difference between the central values of the nominal fit model and the alternative
fit model in which JER variations are implemented coherently
across all bins of the octant variable distribution.
The total systematic uncertainty is calculated as the sum in
quadrature of the individual grouped sources.
}
\end{table}
 
\begin{figure}[!htbp]
\begin{center}
\includegraphics[width=0.55\textwidth]{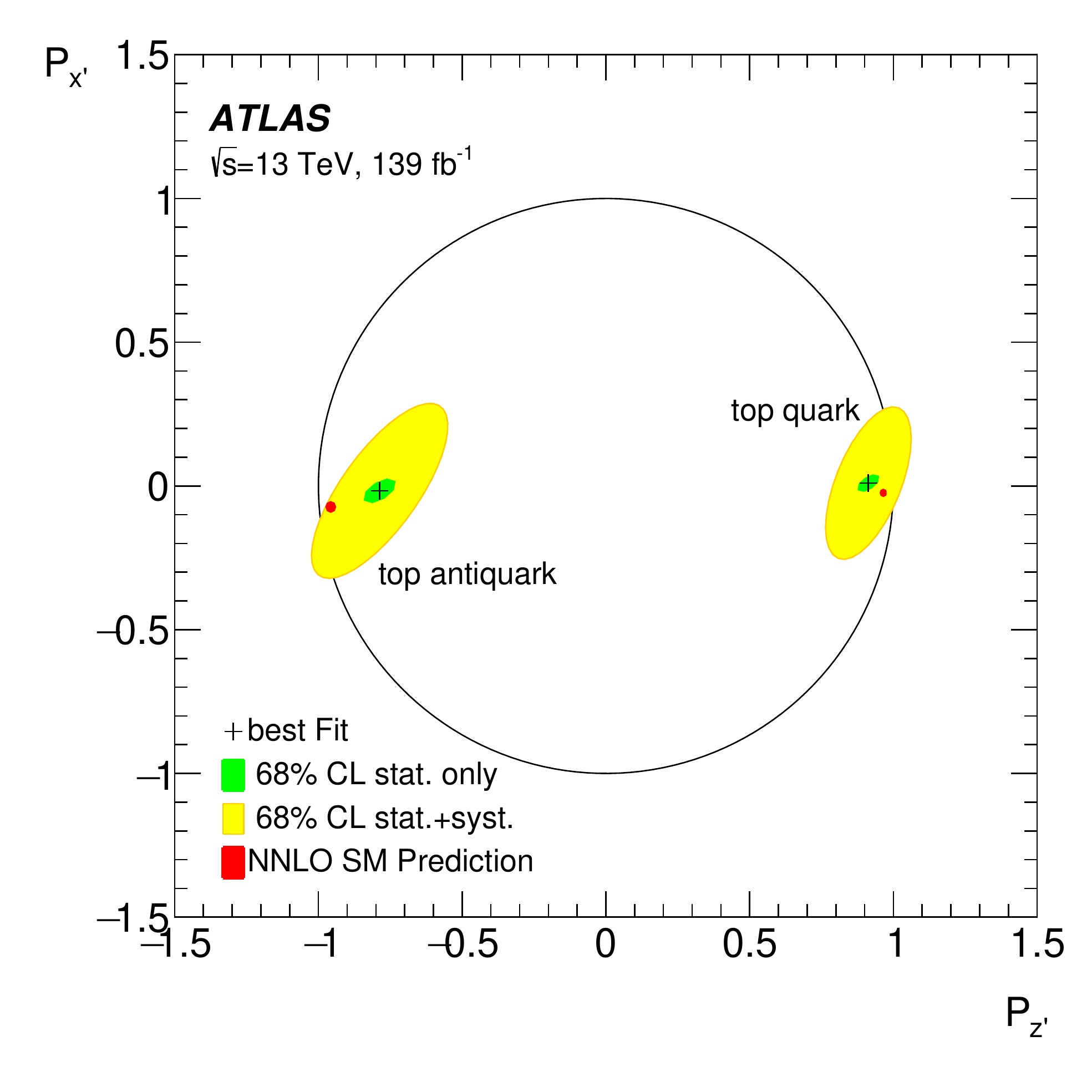}
\caption{Summary of the observed best-fit polarisation measurements with their
statistical-only (green) and statistical+systematic (yellow) contours at 68\% CL,
plotted on the two-dimensional polarisation parameter space
($\Polz,\,\Polx$). The interior of the black circle represents the
physically allowed region of the parameter
space, and the red point indicates
the parton-level prediction at NNLO from a calculation based on Ref.~\cite{Berger:2017zof}.
The uncertainty in the theoretical prediction includes scale, \alphas\ and PDF uncertainties.
Correlations between the predictions of the polarisation parameters are not provided.
\label{fig:elliptical_data}}
\end{center}
\end{figure}

\Figure{fig:elliptical_data} shows the observed best-fit
polarisation measurements of $\Polz$ and $\Polx$ in the
two-dimensional polarisation parameter space. Contours for
top quarks and top antiquarks are shown separately at
68\% CL, including statistical and systematic
uncertainties.
 
From the present analysis one can conclude that a very high degree of
polarisation is observed in \tch\ production,
primarily along the direction of the spectator quark (for top-quark
events), or opposite to that direction (for top-antiquark events),
in agreement with NNLO QCD predictions.


\section{Angular differential cross-sections for top-quark production}
\label{sec:unfolding-method}
 
Three angular differential cross-sections are obtained from the
distribution of  the charged-lepton momentum with respect to the $i^\mathrm{th}$ axis
($i = x', y', z'$), as defined in \Section{sec:axes_definition}. These three
distributions are associated with the three top-quark polarisation
components ($\{\Polx,\Poly,\Polz\}$).
The selection criteria for reconstructed events, defined in \Section{sec:event-selection}, are the same as the used for the
measurement of top-quark and top-antiquark polarisation vectors
in \Section{sec:template-fit}.
These differential angular distributions are distorted by finite resolution of the detector and the trigger,
reconstruction, and sculpted by the
event selection criteria.
The detector effects are corrected using an unfolding technique.
 
In the measurement of the angular differential cross-sections, the
normalisation of the \Wjets and top-quark
background (i.e.\ \ttbar, $\Wt$, and \sch) contributions is estimated through a
simultaneous maximum-likelihood fit to the numbers of data events
observed in the signal region and the two control regions.
All other backgrounds are fixed to their theoretical or data-driven predictions.
The overall normalisation of the \tch\ signal is treated as another free parameter in the fit.
The likelihood function~\cite{TOPQ-2011-14} is given by the product of Poisson
probability terms associated with the three regions (signal region and
\ttbar\ and \Wjets\ control regions).
The fit is performed separately for the signal and control regions
defined with positively (negatively) charged leptons for the
measurement of top-quark (top-antiquark) events, as well as for the
inclusive regions containing both lepton charges.
These normalisation factors are compatible with those
extracted in the template fit described in \Section{sec:template-fit}
and with the theoretical predictions.

\begin{figure}[!htb]
\begin{center}
\subcaptionbox{}{\includegraphics[width=0.42\textwidth]{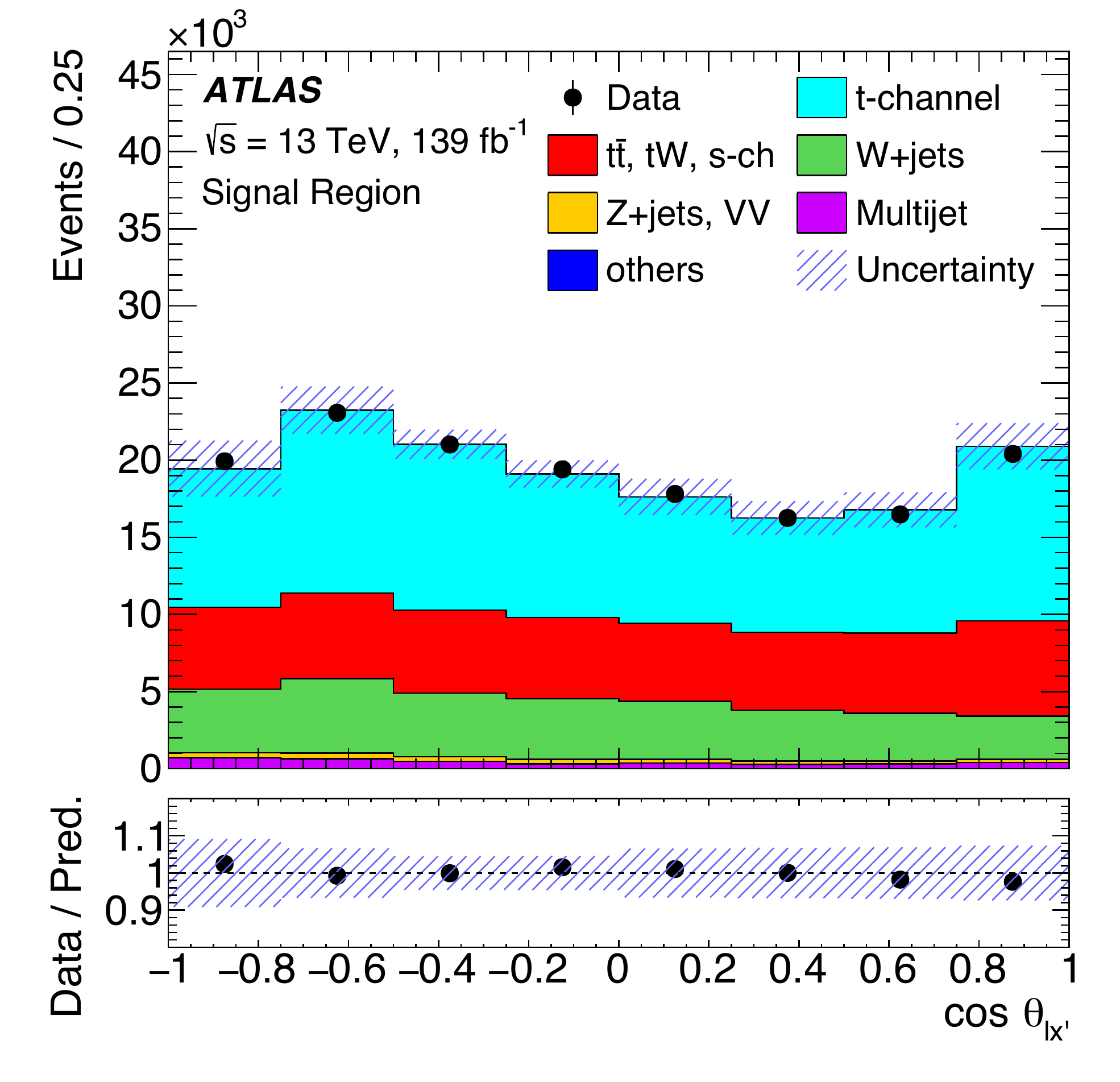}}
\subcaptionbox{}{\includegraphics[width=0.42\textwidth]{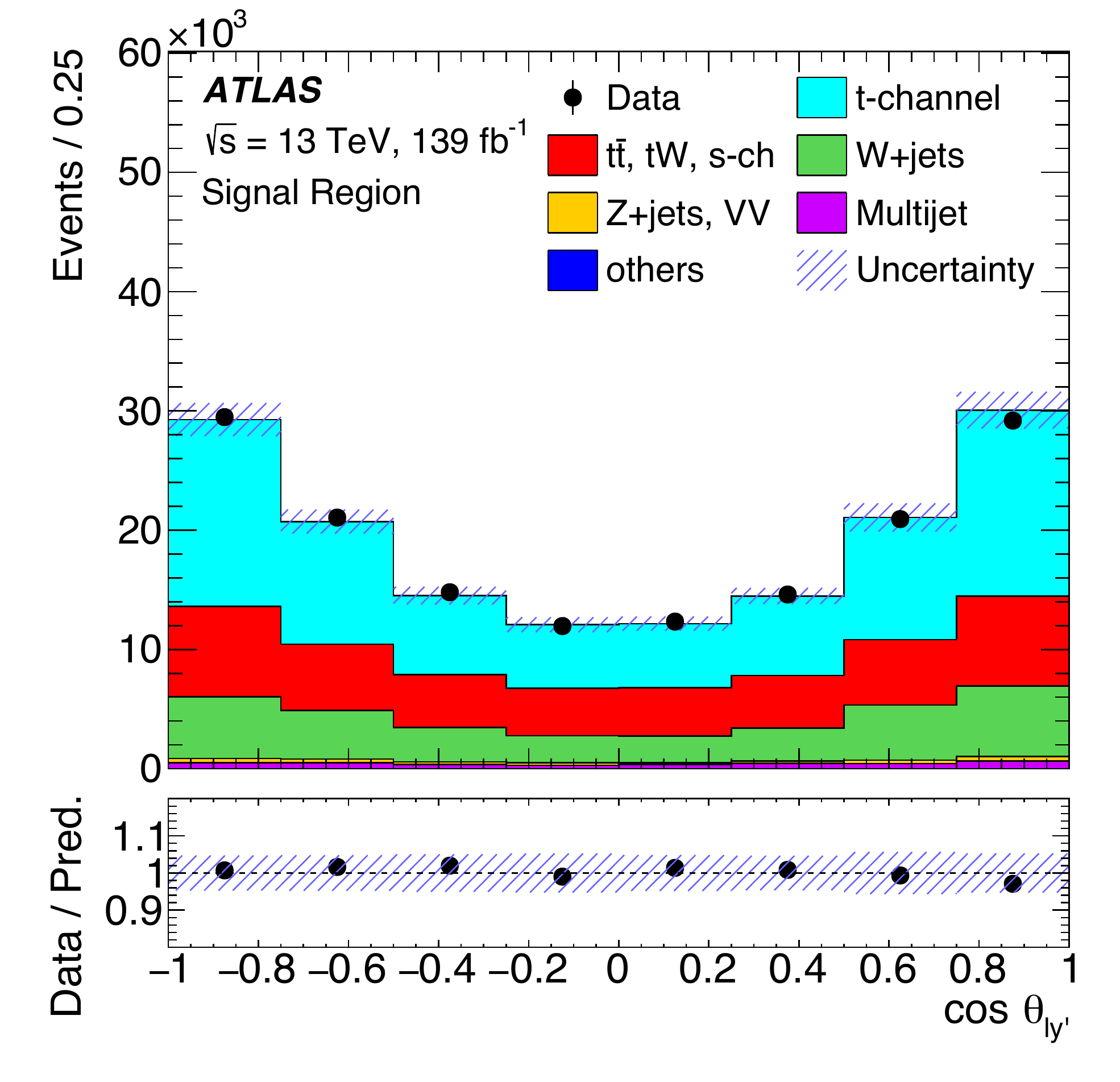}}
\subcaptionbox{}{\includegraphics[width=0.42\textwidth]{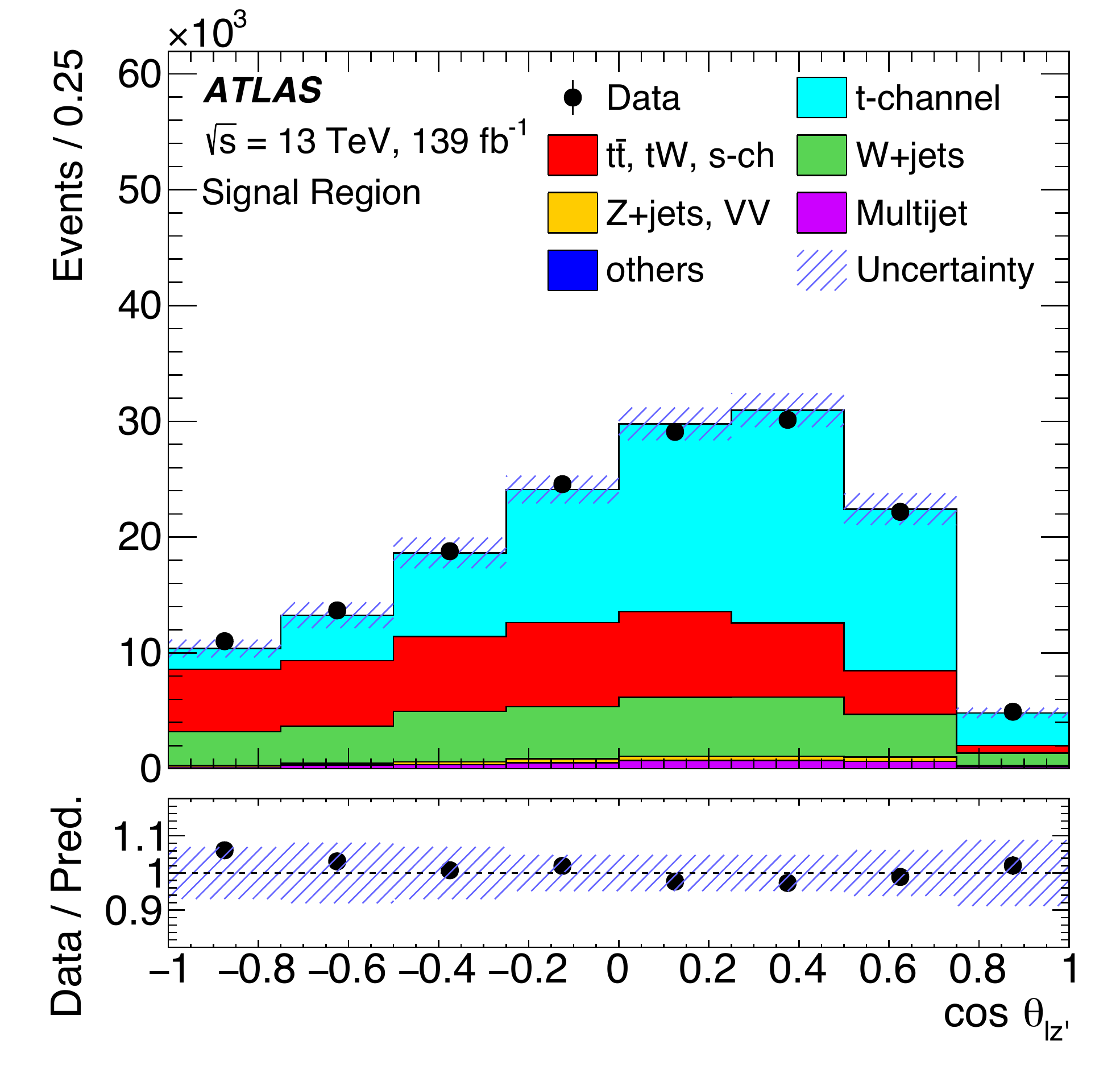}}
\caption{Post-fit distributions of (a) \costhetax, (b) \costhetay
and (c) \costhetaz in the signal region.
The data, shown as the black points with statistical
uncertainties, are compared with SM signal and background predictions. The \qcd\
background is estimated using MC and data-driven techniques, while
contributions from simulated \Wjets and top-quark background and
\tch\ signal event samples are normalised to the results of a
maximum-likelihood fit to event yields in the signal and control
regions. The label `others' represents \ttZ, \ttW, \tZq, \tH, and \tWZ\ production.
The uncertainty bands include both the statistical
and systematic uncertainties. The lower panels show the ratio of
data to prediction in each bin.}
\label{fig:srangdist_sel_postFit}
\end{center}
\end{figure}

The angular distributions observed at reconstruction level
are shown in \Figure{fig:srangdist_sel_postFit} for the inclusive signal region.
They are compared with the
predicted signal and background distributions, normalised to the
results of the maximum-likelihood fit. The selection requirements
have a significant impact on the shapes of these distributions.

The measured angular distributions are unfolded to the particle level
within the fiducial region. The particle-level selection criteria are discussed
in \Section{sec:particle-selection}. The unfolding corrections account for
distortions due to detector resolution and efficiencies so as to allow
direct comparison with theoretical predictions.
In this analysis the same unfolding technique
as in Ref.~\cite{Aaboud:2017pdi} is used.
D'Agostini's iterative Bayesian approach~\cite{DAgostini:1994fjx}
as implemented in \texttt{RooUnfold}~\cite{Adye:2011gm} is
used to unfold the distributions.
The measured expectation value for the number of signal
events at particle level in each bin $k$ of the fiducial volume, $\nu^{\text{particle}}_{k}$, is obtained from the
observed number of events in each bin $j$ of the reconstructed
distribution $N^{\text{data}}_{j}$, after
subtracting the sum of all background contributions $B_{j}$, according to
 
\begin{equation*}
\nu^{\text{particle}}_{k} = C^{\text{particle!reco}}_{k} \sum_{j} M_{jk}^{-1}C_{j}^{\text{reco!particle}}(N^{\text{data}}_{j} - B_{j}),
\end{equation*}
 
where $M_{jk}$ is the migration matrix which relates the particle-level and
reconstructed values, and $C^{\text{reco!particle}}_{k}$ is a correction
factor that accounts for events that pass reconstruction selection but
not particle-level selection. It is defined as
 
\begin{equation*}
C^{\text{reco!particle}}_{j} = \frac{S^{\text{reco}}_{j} - S^{\text{reco!particle}}_{j}}{S^{\text{reco}}_{j}},
\end{equation*}
where $S^{\text{reco}}_{j}$ is the number of reconstructed signal events in bin $j$ and $S^{\text{reco!particle}}_{j}$
is the number of events that pass the reconstruction-level selection
but not the particle-level selection.
Another correction factor, $C^{\text{particle!reco}}_{j}$, accounts for
signal events that pass the particle-level selection but not the reconstruction-level selection:
 
\begin{equation*}
C^{\text{particle!reco}}_{k} = \frac{1}{\epsilon_{k}} = \frac{S^{\text{particle}}_{k}}{S^{\text{particle}}_{k} - S^{\text{particle!reco}}_{k}},
\end{equation*}
 
where $S^{\text{particle}}_{k}$ is the number of signal events at particle level and $S^{\text{particle!reco}}_{k}$
is the number of events that pass the particle-level selection but not
the reconstruction-level selection. The factor
$C^{\text{particle!reco}}_{k} $ is the inverse of the efficiency
$\epsilon_{k}$ for signal events at particle level in bin $k$ to pass
the reconstruction-level requirements.
 
Each normalised differential cross-section is determined by dividing
the obtained $\nu^{\text{particle}}_{k}$ value in each bin by the
integral over all bins.

The migration matrix and the selection efficiency are computed using samples of \tch\ signal events
simulated with the \POWBOXPYTHIA8 generator described in \Section{sec:data_mc_samples}.
They are calculated for the signal region defined with positively (negatively) charged leptons for the measurement of top-quark (top-antiquark) events, as well as for the inclusive signal region containing both lepton charges.
The obtained $C^{\text{reco!particle}}_{k}$ correction factors
are around 50\% and the efficiencies $\epsilon_{k}$ around 20\%
for all bins.
 
The number of bins was chosen in order to have a stable unfolding
response with at least 70\% of the events in the diagonal elements of
the migration matrix.
This criterion results in eight bins for the \costhetax\ and \costhetay\ distributions and four bins for the \costhetaz\ distribution for which larger migrations are observed.
The number of iterations is chosen such
that the absolute change between two successive steps becomes
negligible (the difference must be smaller than 0.1\% of the integral of the associated distribution) and when a convergent
state is reached the difference between the results
obtained with the chosen number of iterations and with 15 more
iterations should not exceed 0.15\% of the integral of the associated distribution for all bins. This stability criterion
results in five iterations for the \costhetax distribution and three
for both the \costhetay and \costhetaz distributions.
 
The stability of the unfolding procedure was
validated through convergence and closure tests performed by using
template distributions constructed from the \tch\
\POWBOXPYTHIA8, \PROTOSPYTHIA8 and $\MGNLOPY8$ samples. The
closure tests showed that the residual bias induced by the unfolding
method is negligible. By using template distributions given by the
\PROTOSPYTHIA8 and $\MGNLOPY8$ samples, generated 
including effects on the $\tWb$ vertex from anomalous couplings or
additional EFT operators, it is shown
that the unfolding method recovers the generated distributions at
particle level within the fiducial region.
 
The statistical uncertainty of the data unfolded result is determined
by running over an ensemble of 100\,000 pseudo-experiments, varying the content
of each bin according to its expected statistical uncertainty through
Poisson fluctuations. For each pseudo-experiment a new background
normalisation is extracted using the procedure described above. After
background subtraction, each pseudo-experiment is unfolded and normalised. The
spread (RMS) of the result in each bin is taken as the measure of the
statistical uncertainty.

The uncertainty associated with the finite size of the simulated samples
is evaluated using 100\,000 pseudo-experiments, varying for each
process (excluding the data-driven \qcd\ background) the total prediction in each bin according to the
statistical uncertainty through Gaussian fluctuations. The same
procedure as for the data statistical uncertainty is then used.
 
The impact of each source of
systematic uncertainty is evaluated by unfolding template
distributions resulting from simulated pseudo-data modified to reflect the effect of the given uncertainty source.
In each case a new background normalisation estimation is
performed before its subtraction from the pseudo-data, using the
fitting procedure described above. For all sources of systematic uncertainty the nominal
unfolding corrections are considered. Except for the signal modelling
uncertainties affecting the unfolding corrections, the systematic
uncertainty is evaluated as the difference between the nominal angular
distribution values and the ones measured using the varied
normalisations and shapes. For the signal modelling uncertainties, the
unfolded varied distribution is compared with the corresponding
particle-level spectra.

\begin{figure}[!htb]
\begin{center}
\subcaptionbox{}{\includegraphics[width=0.45\textwidth]{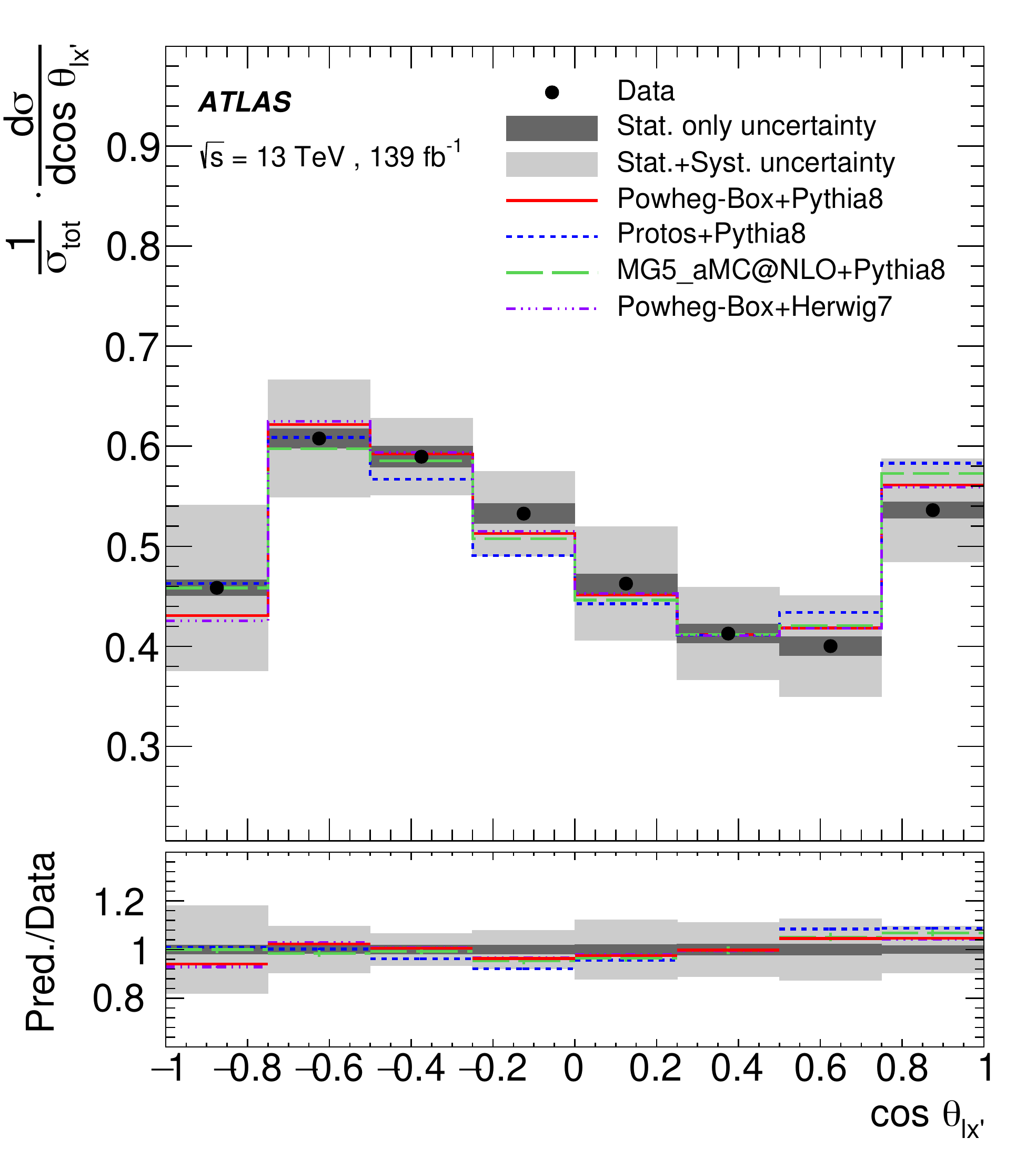}}
\subcaptionbox{}{\includegraphics[width=0.45\textwidth]{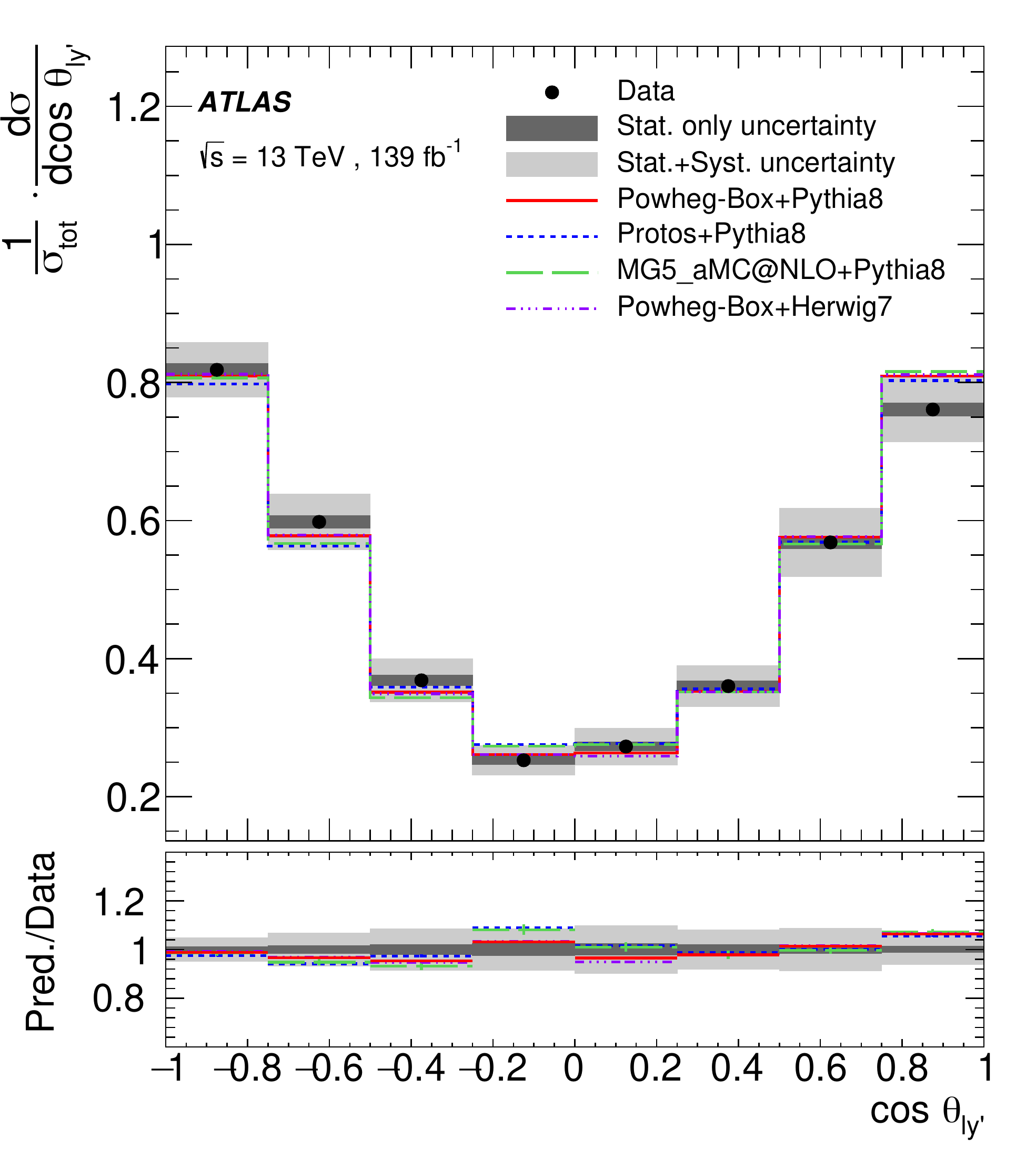}}
\subcaptionbox{}{\includegraphics[width=0.45\textwidth]{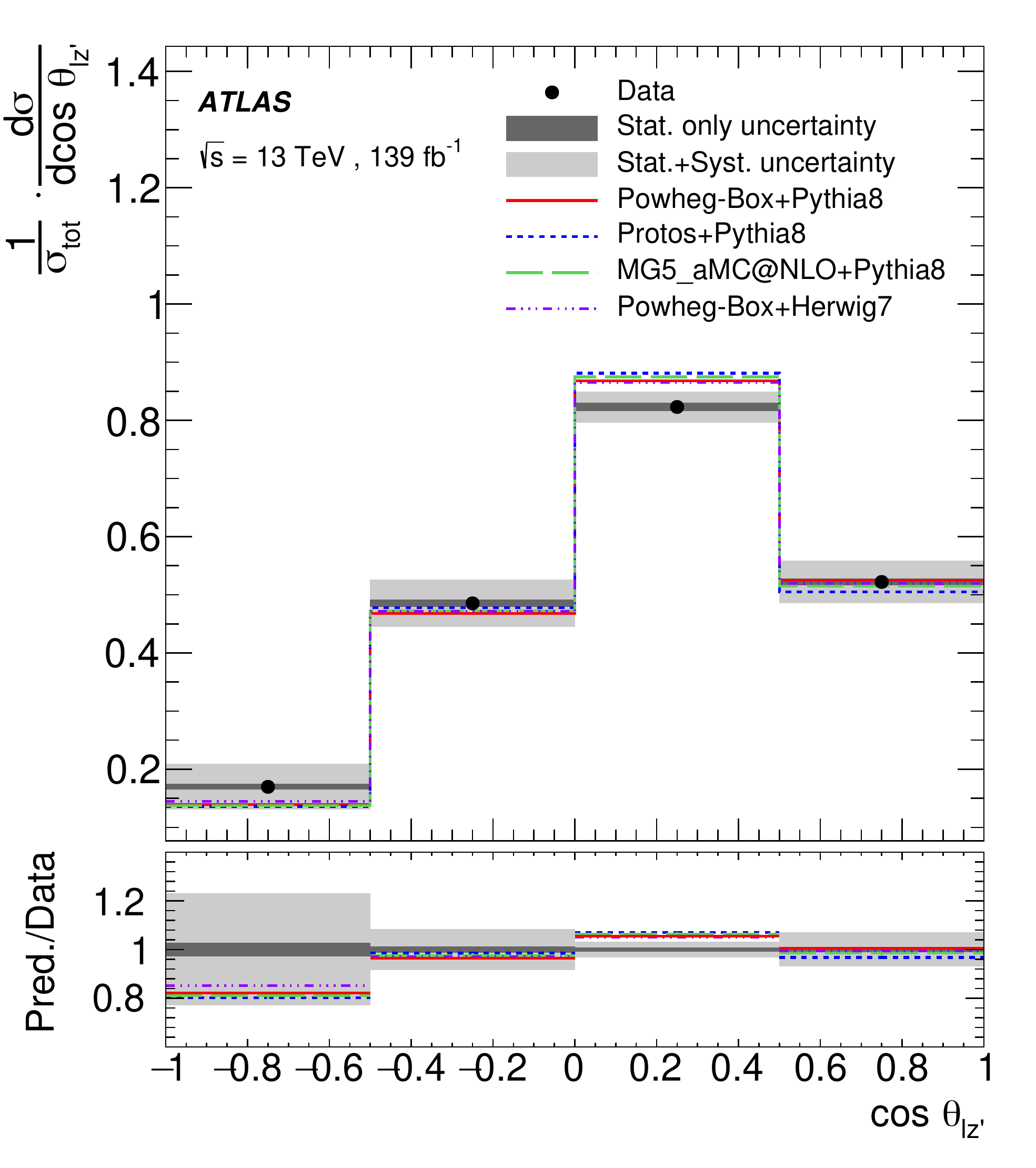}}
\caption{Particle-level normalised differential cross-sections as a function of  (a)
\costhetax, (b) \costhetay, and (c) \costhetaz, along with
various SM MC predictions of the \tch\ signal for both top quarks
and top antiquarks.
The data, shown as the black points with statistical
uncertainties, are compared with predictions (lines) obtained by using the
\POWBOXPYTHIA8\ (solid red), \PROTOSPYTHIA8\ (dashed blue),
$\MGNLOPY8$ (long-dashed green) and
\POWBOXHERWIG7\ (dot-dashed violet) generators. The uncertainty
bands include both the statistical
and systematic uncertainties. The data statistical uncertainty is too small to be visible. The lower panels show the ratio of prediction to
data in each bin.}
\label{fig:UnfoldINC}
\end{center}
\end{figure}

The resulting normalised angular differential cross-sections at particle level within the fiducial
region are displayed in \Figures{fig:UnfoldINC}{fig:UnfoldANTITOP} for the inclusive, top-quark and top-antiquark measurements. The measured cross-sections include both the statistical and
systematic uncertainties, and are compared with the predictions given by the
different generators. Among the different uncertainty sources, those related
to JER and JES are dominant.
The measured distributions are consistent with those obtained
from a \PROTOS\ simulation sample with the polarisations given by the
template fit described in \Section{sec:template-fit}.
 
\begin{figure}[!htb]
\begin{center}
\subcaptionbox{}{\includegraphics[width=0.45\textwidth]{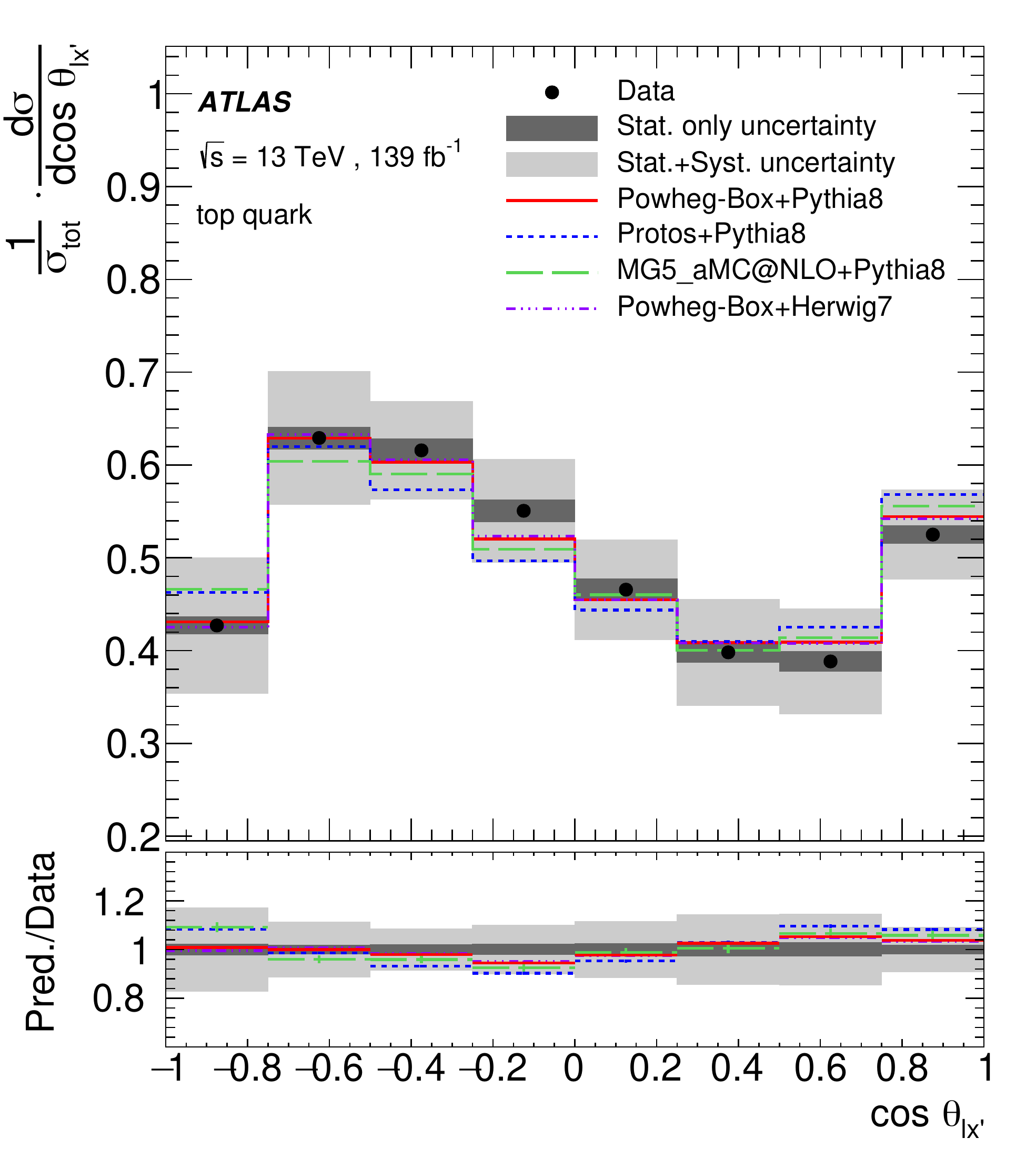}}
\subcaptionbox{}{\includegraphics[width=0.45\textwidth]{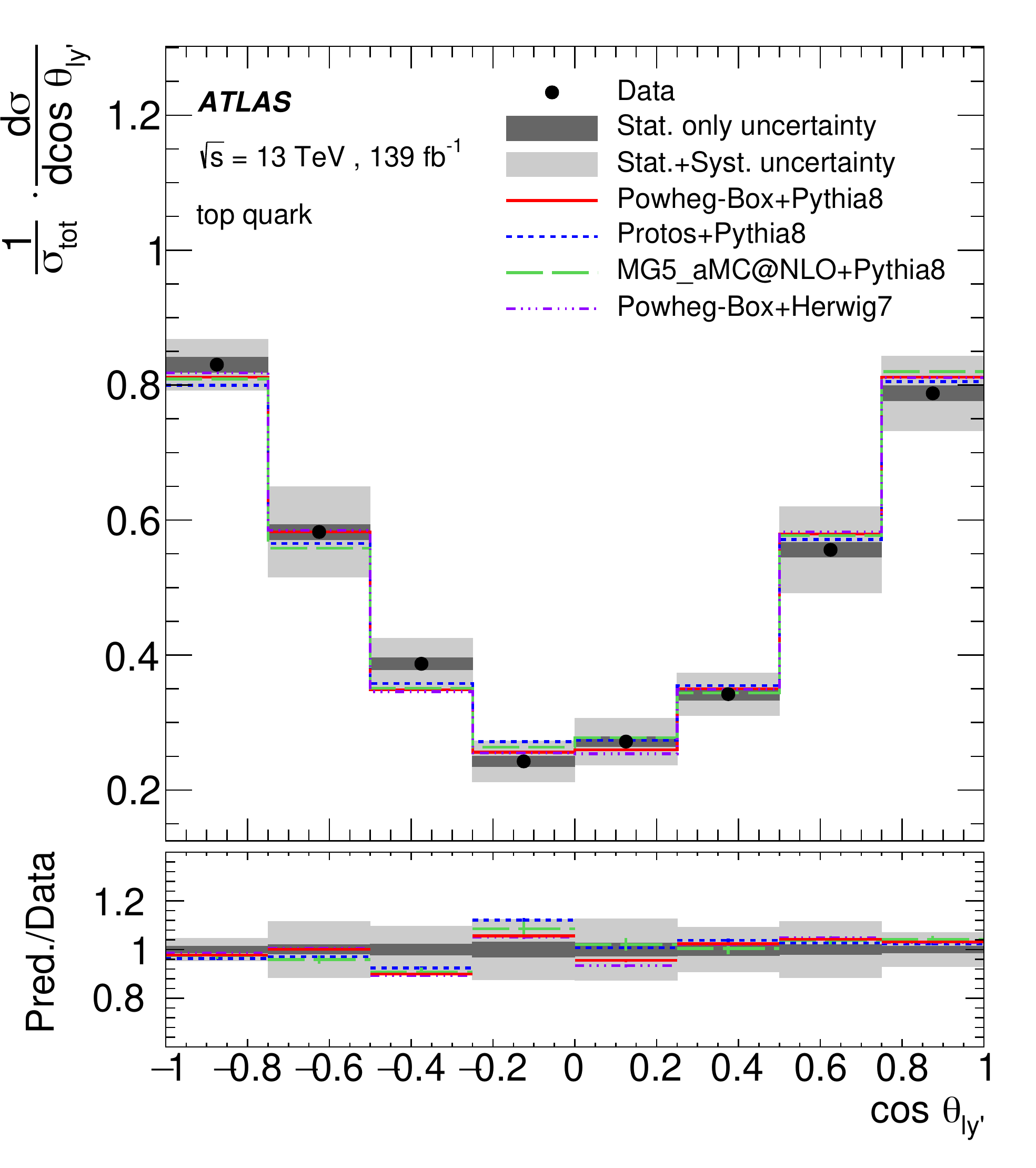}}
\subcaptionbox{}{\includegraphics[width=0.45\textwidth]{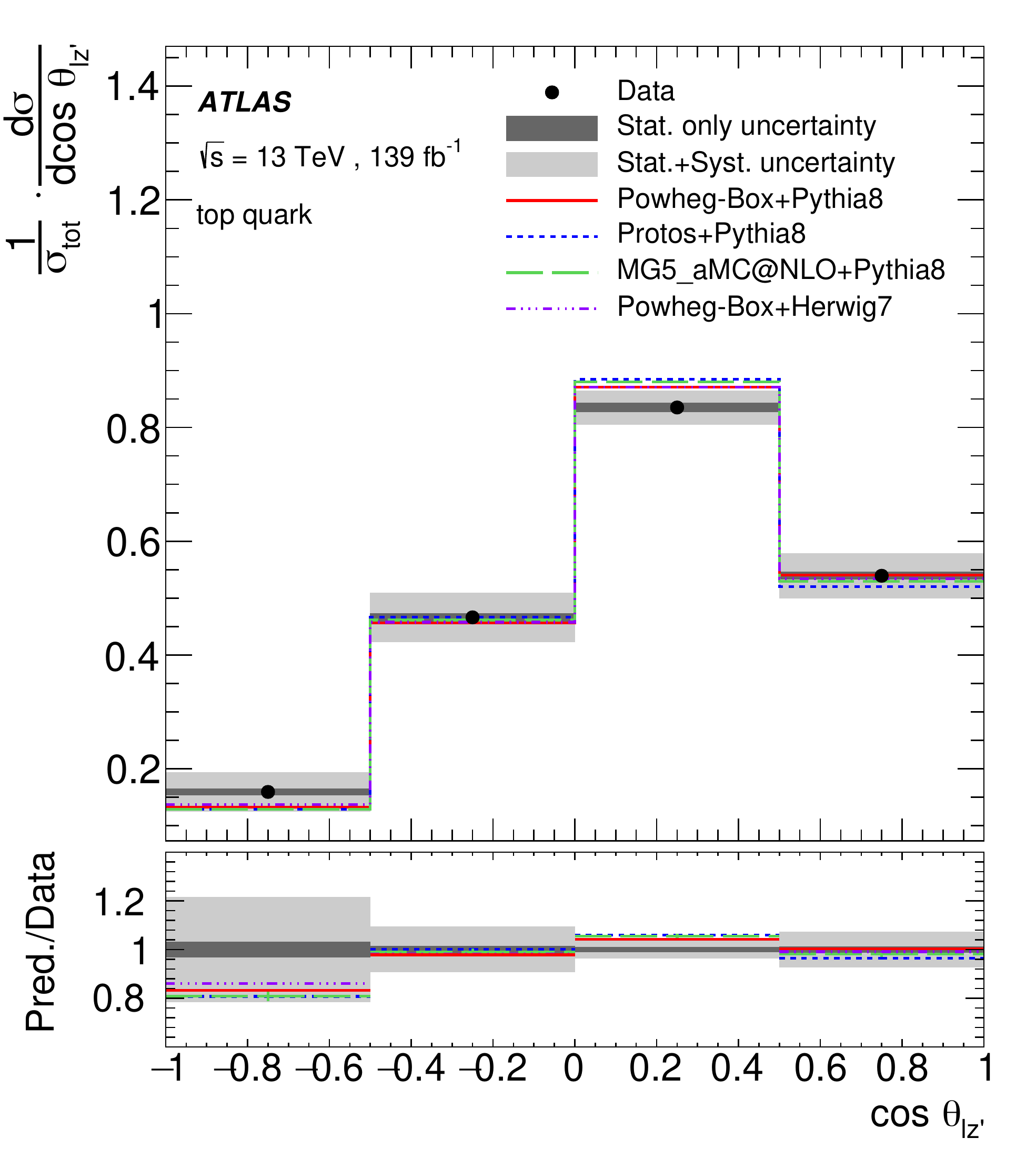}}
\caption{Particle-level normalised differential cross-sections as a function of  (a)
\costhetax, (b) \costhetay, and (c) \costhetaz, along with
various SM MC predictions of the \tch\ signal for top quarks.
The data, shown as the black points with statistical
uncertainties, are compared with predictions (lines) obtained by using the
\POWBOXPYTHIA8\ (solid red), \PROTOSPYTHIA8\ (dashed blue),
$\MGNLOPY8$ (long-dashed green) and
\POWBOXHERWIG7\ (dot-dashed violet) generators. The uncertainty
bands include both the statistical
and systematic uncertainties. The data statistical uncertainty is too small to be visible. The lower panels show the ratio of prediction to
data in each bin.}
\label{fig:UnfoldTOP}
\end{center}
\end{figure}
 
\begin{figure}[!htb]
\begin{center}
\subcaptionbox{}{\includegraphics[width=0.45\textwidth]{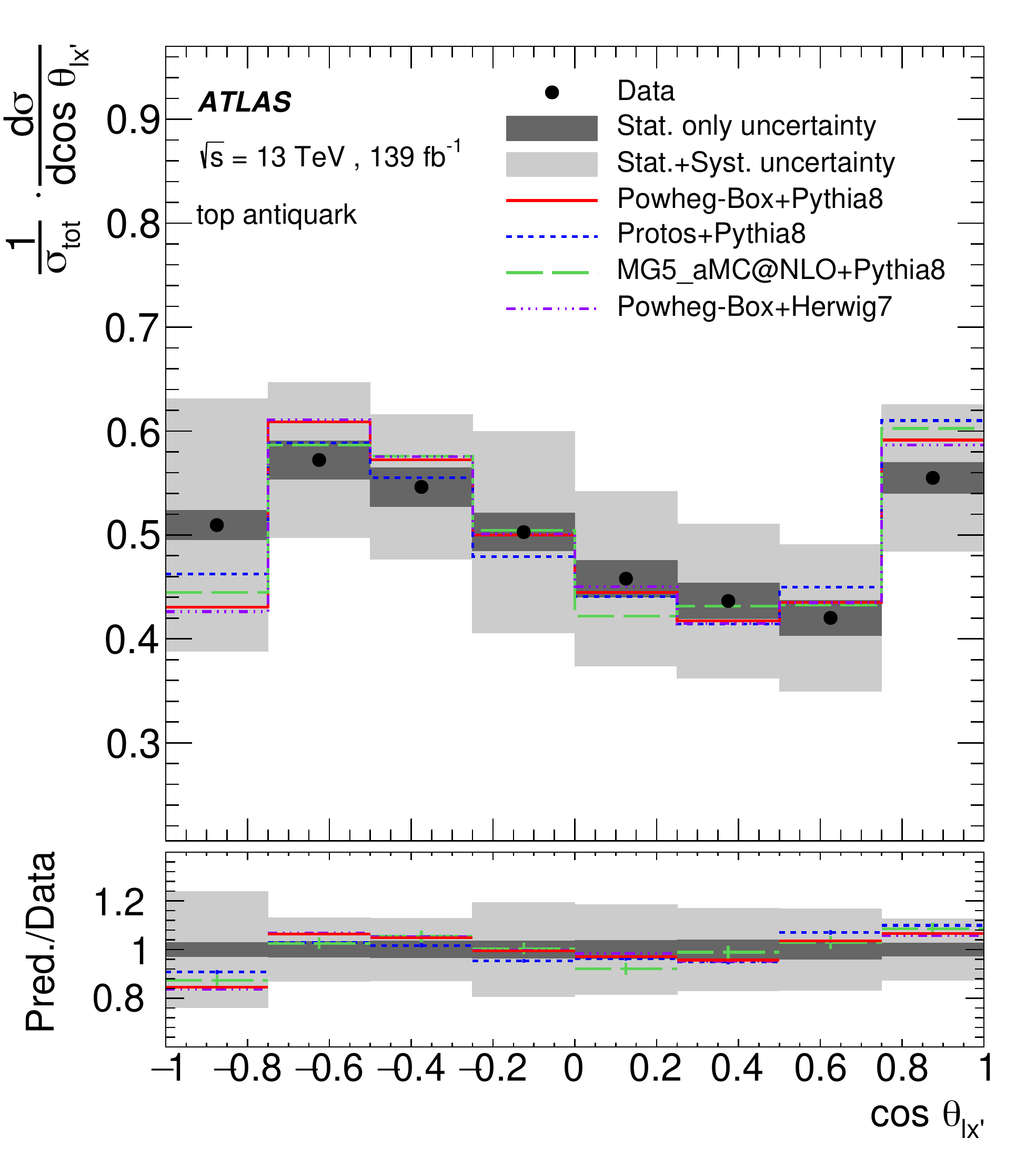}}
\subcaptionbox{}{\includegraphics[width=0.45\textwidth]{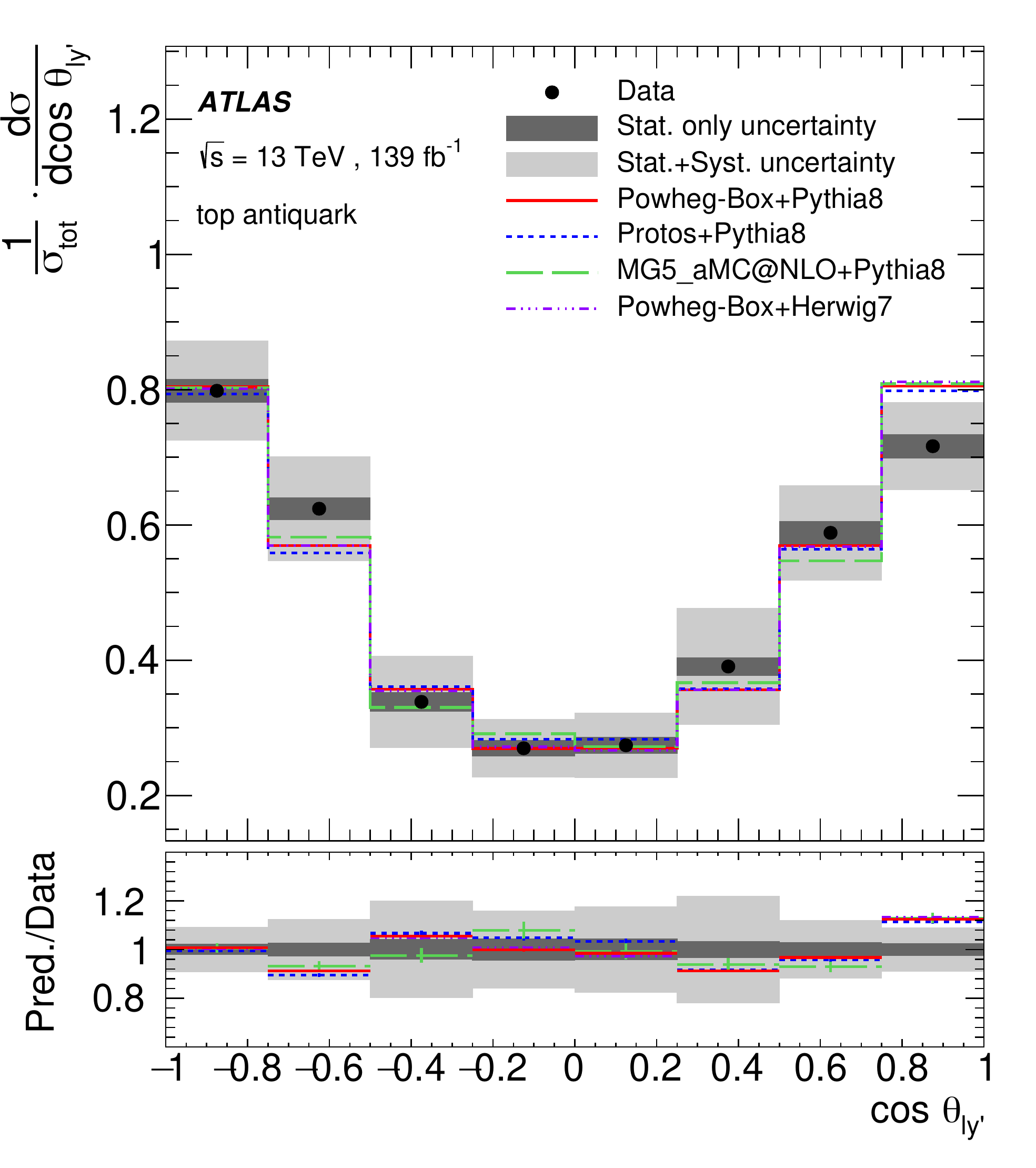}}
\subcaptionbox{}{\includegraphics[width=0.45\textwidth]{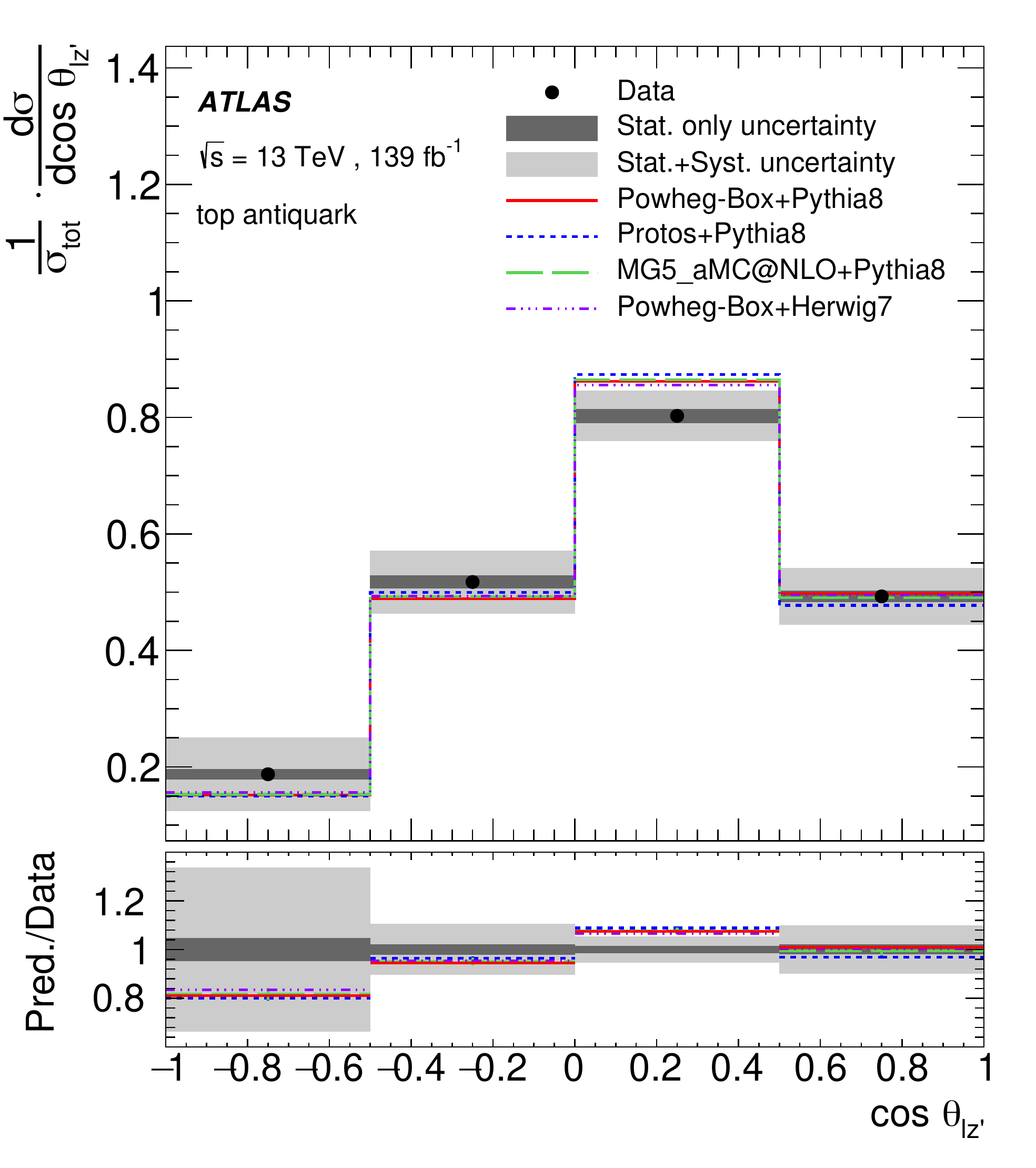}}
\caption{Particle-level normalised differential cross-sections as a function of  (a)
\costhetax, (b) \costhetay, and (c) \costhetaz, along with
various SM MC predictions of the \tch\ signal for top antiquarks.
The data, shown as the black points with statistical
uncertainties, are compared with predictions (lines) obtained by using the
\POWBOXPYTHIA8\ (solid red), \PROTOSPYTHIA8\ (dashed blue),
$\MGNLOPY8$ (long-dashed green) and
\POWBOXHERWIG7\ (dot-dashed violet) generators. The uncertainty
bands include both the statistical
and systematic uncertainties. The data statistical uncertainty is too small to be visible. The lower panels show the ratio of prediction to
data in each bin.}
\label{fig:UnfoldANTITOP}
\end{center}
\end{figure}
 
A global covariance matrix of size 20$\times$20, corresponding to the 20 measured bins in \Figures{fig:UnfoldINC}{fig:UnfoldANTITOP} and including the effects of all uncertainties and the correlations among the three angular distributions, is computed in order to perform quantitative comparisons with theoretical predictions.
 
An initial ensemble of 50\,000 pseudo-experiments is used to compute the covariance matrix with the systematic uncertainties and the uncertainty associated with the limited size of the simulated samples. In
each pseudo-experiment, Gaussian-distributed shifts are applied
coherently for each detector-modelling systematic uncertainty by
scaling each bin of the data distribution by the expected relative
variation from the associated systematic uncertainty effect. For the uncertainty due to the limited size of the simulated samples the Gaussian-distributed shifts are computed independently for each bin. The
varied distribution is unfolded with the nominal corrections after subtracting the backgrounds with the normalisation factors
derived following the procedure described in \Section{sec:background-estimation}.
Additional Gaussian-distributed shifts are then applied coherently for each
signal- and background-modelling systematic uncertainty. The modelling
shifts are derived by using the expected relative variations from the
associated systematic uncertainty to scale each bin of the
data distribution unfolded with the nominal corrections. The
resulting modified unfolded distributions are used to compute an initial matrix.
 
The statistical correlations between the three different angular distributions are evaluated using the bootstrap method~\cite{Bohm:389738}, using a set of 1000 bootstrap samples. The global statistical matrix is calculated using the statistical error of each bin of the angular distributions and the correlation coefficients. This matrix is then added to the previous one to compute an unnormalised global covariance matrix.
 
Finally, a set of 100\,000 additional pseudo-experiments is used to normalise the covariance matrix. For each additional pseudo-experiment, each element of the three angular distributions is fluctuated using a multivariate Gaussian distribution with the unnormalised global covariance matrix. The resulting pseudo-experiments are normalised and are used to calculate the normalised global covariance matrix.
 
To quantify the level of agreement between each of the measured normalised differential cross-sections and the theoretical
predictions, $\chi^{2}$ values are calculated according to the relation
$\chi^{2} = V_{N_{b}-1}^{T} \cdot \text{Cov}_{N_{b}-1}^{-1} \cdot V_{N_{b}-1}$.
Here $N_{b}$ is the number of bins in the spectrum under
consideration, $V_{N_{b}-1}$ is a vector with the differences between
the measured and the predicted cross-sections obtained by discarding
one of the $N_{b}$ elements,  and $\text{Cov}_{N_{b}-1}$
is the $(N_{b}-1) \times (N_{b}-1)$ sub-matrix derived
from the relevant $N_{b} \times N_{b}$ part of the
global covariance matrix by discarding the corresponding row and column.
A global $\chi^{2}$ value is also obtained using the full vector of differences
for the three angular distributions and the full global covariance matrix.
In this case, three elements of the vector and their corresponding rows and columns
in the covariance matrix are discarded.

The $p$-values, relative to the \POWBOXPYTHIA8 SM predictions, are then evaluated from the $\chi^{2}$ values and the number
of degrees of freedom (NDF) of each angular distribution, as shown in
\Table{tab:chipval}. Since these are normalised
cross-sections, the NDF corresponds to the number of bins in each
angular distribution minus one. The obtained values show good agreement of the
SM prediction with the measured data. The high $p$-values for
$\costhetax$ distribution may indicate that the evaluation of
systematics uncertainties is overly conservative.

\begin{table}[!htb]
\begin{center}
\begin{tabular}{c|cc|cc|cc}
\hline
\hline
& \multicolumn{2}{c|}{Top quark} & \multicolumn{2}{c|}{Top antiquark} & \multicolumn{2}{c}{Top quark and antiquark} \\
\hline
Angular variable & $\chi^{2}$/NDF & $p$-value & $\chi^{2}$/NDF & $p$-value & $\chi^{2}$/NDF & $p$-value \\
\hline
$\costhetax$ & 1.35/7 & 0.99 & 0.94/7 & 1.00 & 1.32/7 & 0.99 \\
$\costhetay$ & 4.57/7 & 0.71 & 2.92/7 & 0.89 & 3.78/7 & 0.81 \\
$\costhetaz$ & 1.55/3 & 0.67 & 2.04/3 & 0.56 & 2.26/3 & 0.52 \\
\hline
Global & 13.55/17 & 0.70 & ~~6.86/17 & 0.99 & ~~9.25/17 & 0.93 \\
\hline
\hline
\end{tabular}
\caption{The $\chi^{2}$ and $p$-value of the three unfolded
angular distributions for the top-quark, for the top-antiquark
and for both the top-quark and top-antiquark measurements. The numbers are computed by comparing the observed data with the \POWBOXPYTHIA8 SM predictions. The NDF corresponds to the number of bins of each angular distribution minus one. A global $\chi^{2}$ and $p$-value for the three angular distributions are also included.}
\label{tab:chipval}
\end{center}
\end{table}


\section{Bounds on EFT coefficients \ctW\ and \citW}
\label{sec:eft_interpretation}
 
The unfolded and normalised distributions of \costhetax\ and \costhetay\ are used to set bounds on 
the complex Wilson coefficient of the dimension-six operator
$\mathcal{O}_{tW}$.
The operator $\mathcal{O}_{tW}$ has only a small effect on the \costhetaz\ distribution, which in contrast to the \costhetax\ and \costhetay\ distributions is sensitive to many additional EFT operators~\cite{Aguilar-Saavedra:2017nik}. The \costhetaz\ distribution is therefore ignored in this work.
The data are compared with the effects of this dipole operator in a likelihood fit which requires a theoretical prediction in a parametric form.
To describe the angular distributions as a function of \ctW\ and \citW, a morphing technique~\cite{Baak:2014fta,ATL-PHYS-PUB-2015-047} is employed to interpolate from a set of 15 MC templates generated as described in \Section{sec:data_mc_samples}.
The $\mathcal{O}_{tW}$ operator contributes to the production and decay of the top quark.
All orders of the EFT expansion parameter $\Lambda$ are included in the parametric description,
allowing studies of the effects from
high-dimensional (up to $1/\Lambda^8$) terms, which are usually assumed negligible.
 
The EFT operator can contribute to both the production and decay of the top quark, which leads to 15 terms in the weight function:
\begin{align*}
\label{eq:morph_full_expression}
\sigma (\ctW, \citW) & \propto  \left | \mathcal{O}_{\text{SM}} + \frac{\ctW}{\Lambda^2} \cdot \mathcal{O}_{tW} + \frac{\citW}{\Lambda^2} \cdot \mathcal{O}_{itW} \right |^2_{\text{production}}  \cdot \left | \mathcal{O}_{\text{SM}} + \frac{\ctW}{\Lambda^2} \cdot \mathcal{O}_{tW} +  \frac{\citW}{\Lambda^2} \cdot \mathcal{O}_{itW} \right |^2_{\text{decay}}  \\
& =    \sigma_1 + \left(\ctW^1 \cdot \sigma_2 + \citW^1 \cdot \sigma_3\right)/\Lambda^2   \\
& +   \left(\ctW^2 \cdot \sigma_4 + \citW^2 \cdot \sigma_5 + \ctW^1\citW^1 \cdot \sigma_6\right)/\Lambda^4 \\
& +  \left(\ctW^3 \cdot \sigma_7 + \citW^3 \cdot \sigma_8 + \ctW^2\citW^1 \cdot \sigma_9 + \ctW^1 \citW^2 \cdot \sigma_{10}\right)/\Lambda^6  \\
& +  \left(\ctW^4 \cdot \sigma_{11} + \citW^4\cdot \sigma_{12} + \ctW^3 \citW^1 \cdot \sigma_{13} + \ctW^1 \citW^3 \cdot \sigma_{14} + \ctW^2 \citW^2 \cdot \sigma_{15}\right)/\Lambda^8.
\end{align*}
This expression includes all orders of the EFT expansion ($\Lambda$)
with the numerical factors originating from the squaring absorbed in
the $\sigma_n$ terms, with $n = 1,...,15$.
The positions in \ctW\ and \citW\ of 15 simulated event samples was chosen such that the uncertainty in the weight function is a small contribution to the total uncertainty over the parameter-space region of interest.
If the Lagrangian had not been truncated at dimension six, dimension-eight operators would also begin to contribute at ($1/\Lambda^4$).
In this work, $\Lambda$ is set to \SI{1}{\TeV}.
 
The likelihood function has the form of a multivariate Gaussian
distribution whose mean is set to the EFT prediction from the measured
angular differential cross-section. The likelihood contains the
uncertainties of the 15 $\sigma_n$ terms, due to the limited size of the MC
samples used to produce the templates, together with the uncertainty of the
unfolded measurement contained in a covariance matrix.
 
The 16$\times$16 covariance matrix for the 16 bins of the measured normalised
\costhetax\ and \costhetay\ distributions is the corresponding sub-matrix of the global matrix described in \Section{sec:unfolding-method}.
This matrix has two redundant dimensions caused by the fact that the
angular distributions \costhetax\ and \costhetay\ are both normalised and are obtained from the same dataset.
In the implementation of the fitting procedure, the covariance matrix
is first diagonalised and the data is projected onto the eigenvectors
with non-zero eigenvalues.
 
The robustness of the fit is tested in three ways.
It was checked that the unfolding procedure which uses
SM-derived corrections can correctly recover non-zero EFT coefficients
over a range that spans the current bounds. Secondly, the effect of
non-zero EFT coefficients on the background subtraction was proven to
be smaller than the measurement uncertainties.
Lastly, when applied to MC samples produced by the
$\MGNLOPY8$ generator with non-zero values for other EFT operator
coefficients (\cbW, \cibW, \cphitb, \ctg), the stability of the fit for \ctW\ and \citW\
was found to be well within the current experimental precision.
 
\begin{figure}[!htb]
\begin{center}
\subcaptionbox{}{\includegraphics[width=0.49\textwidth]{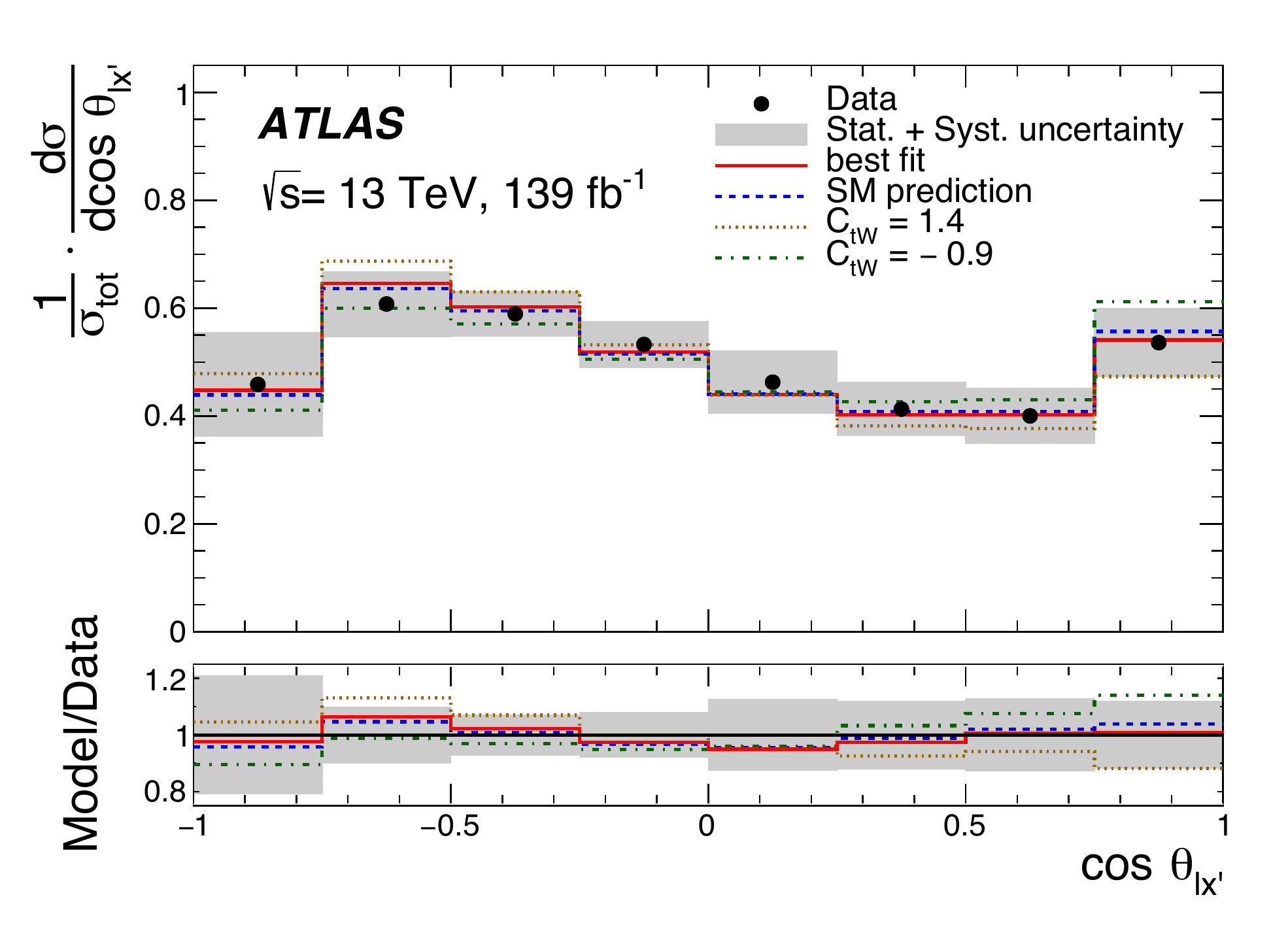}}
\subcaptionbox{}{\includegraphics[width=0.49\textwidth]{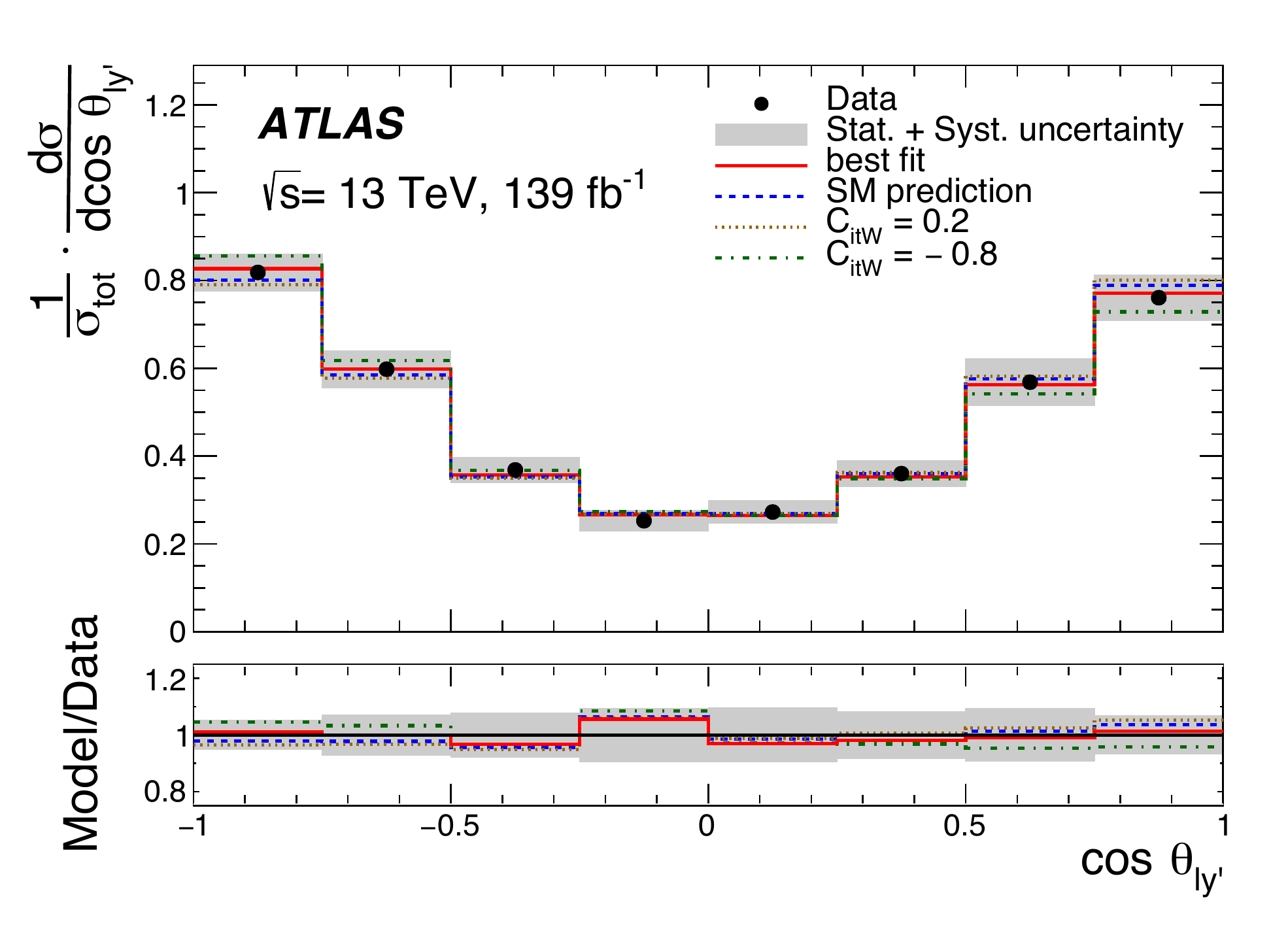}}
\caption{Comparison of data and the result of the EFT fit
for the polarisation angles (a) \costhetax\ and (b) \costhetay.
The solid points show the data, unfolded to particle level.
The solid red line corresponds to the EFT prediction using the
best-fit values for the Wilson coefficients \ctW\ = 0.3 and \citW\ = $-0.3$.
The blue dashed line represents the SM prediction obtained with
the $\MGNLOPY8$ generator.
The brown dotted (green dash-dotted) line shows the model at its
upper (lower) 95\% CL bounds for (a) $\ctW = 1.4$ ($\ctW =
-0.9$) and (b) $\citW = 0.2$ ($\citW = -0.8$) also obtained with the
$\MGNLOPY8$ generator.
The uncertainty bands include both the statistical and systematic uncertainties.
The lower panel gives the ratio of the model to the data.}
\label{fig:eft_result}
\end{center}
\end{figure}

\begin{figure}[!htb]
\begin{center}
\includegraphics[width=0.49\textwidth]{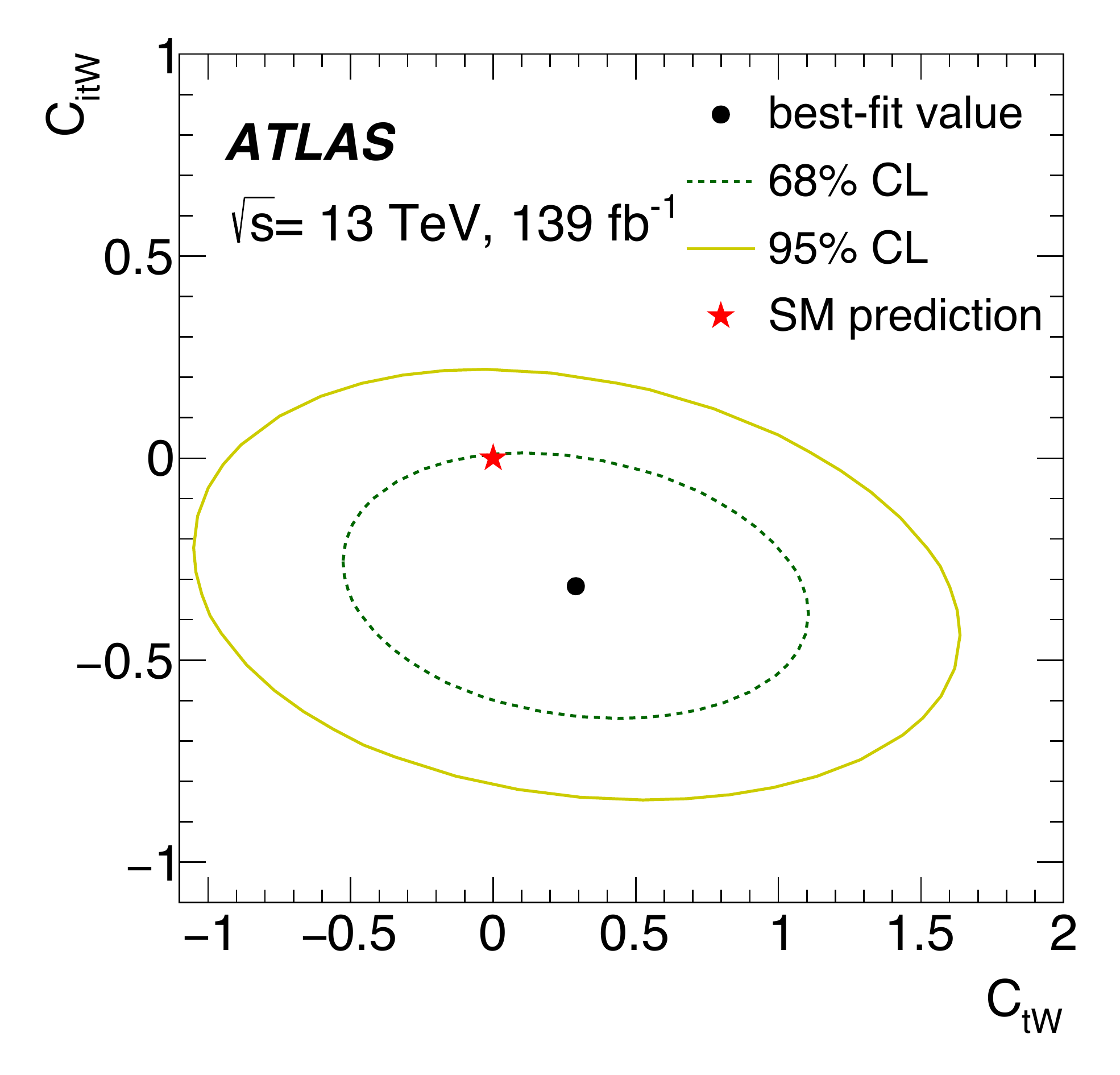}
\end{center}
\caption{The observed best-fit value (dot) for the Wilson
coefficients \ctW\ and \citW\ with the uncertainty contours at 68\% CL
(dashed) and 95\% CL (solid). The CLs are obtained in a
simultaneous fit of the two parameters
using the prediction obtained with the $\MGNLOPY8$ generator.
The red star indicates the SM prediction.}
\label{fig:eft_elipse}
\end{figure}

The result of fitting \ctW\ and \citW\ simultaneously is shown
in \Figure{fig:eft_result}, where good agreement between the model
and the data is observed. Good agreement with the SM prediction from
the NLO $\MGNLOPY8$ generator is also seen.
The best-fit values for the coefficients are $\ctW = 0.3 \pm
0.6~(1.1)$ and $\citW = -0.3 \pm 0.2~(0.5)$ at 68\% CL (95\% CL), which is consistent with the SM prediction, as shown in \Figure{fig:eft_elipse}.
The obtained limits for the coefficients at 68\% CL and 95\% CL are given in \Table{tab:eft_limits}.
The individual fit result for each coefficient is also obtained by
fixing the other coefficient to zero, which results in $\ctW = 0.1
\pm 0.5~(1.1)$ and $\citW = -0.3 \pm 0.2~(0.5)$ respectively, where the
results are quoted for 68\% CL (95\% CL).
 
\begin{table}[!htb]
\centering
\begin{tabular}{l|cc|cc}
\hline
\hline
& \multicolumn{2}{c|}{$\bm{\ctW}$} & \multicolumn{2}{c}{$\bm{\citW}$}\\
& \textbf{68\% CL} & \textbf{95\% CL} & \textbf{68\% CL} & \textbf{95\% CL} \\
\hline
All terms             & $[-0.3, 0.8]$      & $[-0.9, 1.4]$      & $[-0.5, -0.1]$     & $[-0.8, 0.2]$      \\
Order $1/\Lambda^{4}$ & $[-0.3, 0.8]$      & $[-0.9, 1.4]$      & $[-0.5, -0.1]$     & $[-0.8, 0.2]$      \\
Order $1/\Lambda^{2}$ & $[-0.3, 0.8]$      & $[-0.8, 1.5]$      & $[-0.6, -0.1]$     & $[-0.8, 0.2]$      \\
\hline
\hline
\end{tabular}
\caption{Obtained limits on the real (\ctW) and imaginary (\citW) coefficient of the $\mathcal{O}_{tW}$ operator. Also shown are the limits when only the terms up to a specific order in $\Lambda$ are taken into account.}
\label{tab:eft_limits}
\end{table}
 
Among previous constraints on \ctW, the strongest come from measurements of the values of the $W$-boson helicity fractions in top-quark pair decays by ATLAS~\cite{TOPQ-2016-02} and CMS~\cite{CMS-TOP-13-008}, with the combined result~\cite{TOPQ-2018-02} providing bounds of $[-0.48,0.29]$ at 95\% CL when only \ctW\ is allowed to vary.
Comparable limits were also derived from EFT fits of the top-quark sector~\cite{Brivio:2019ius,Hartland:2019bjb,Durieux:2019rbz,ethier2021combined} that include those measurements.
However, the constraints obtained in those analyses assume $\citW=0$, which is not assumed in the result presented here.
Very stringent individual limits on \citW\ were obtained
from electric dipole moment analyses~\cite{Cirigliano:2016nyn} when only \citW\ is allowed to be non-zero, but these become much weaker when multiple EFT coefficients are allowed to vary simultaneously.
In that case, the previous limits on \citW\ were dominated by the input from ATLAS~\cite{TOPQ-2013-12}, which provides bounds on \citW\ of $[-2.3,3.0]$ at 95\% CL.
A global fit including real and imaginary parts of all \Wtb\ operators~\cite{Deliot2018} gives $\ctW \in [-0.8,0.7]$ and $\citW \in [-2.3,1.6]$, where the bound on \citW\ is dominated by the ATLAS result~\cite{TOPQ-2014-17}.
A later ATLAS result~\cite{TOPQ-2015-11} improved the 95\% CL interval to $\citW \in [-0.8,0.7]$.
The bounds on \citW\ presented in this paper improve on all of these results.
 
Also given in \Table{tab:eft_limits} are the limits when a certain cut-off on the order of
$\Lambda$ is applied. It is observed that including terms up to
$1/\Lambda^{4}$, which corresponds to the squared terms
of a dimension-six coefficient, yields identical limits to retaining all 15 terms. The relatively small change in the
result obtained with only $1/\Lambda^{2}$ terms indicates that there is only a modest dependence on $1/\Lambda^{4}$ terms.
The usual assumption that terms beyond $1/\Lambda^{4}$ can be excluded is also validated.

\FloatBarrier
\section{Conclusions}
\label{sec:conclusions}
 
The polarisation of single top quarks and antiquarks produced in the \tch\ has been measured in \lumi\ of \SI{13}{TeV} $pp$ collision data collected with the ATLAS detector at the LHC.
An analysis of the full polarisation vector of the top quarks
(antiquarks) finds very high polarisation along (against) the
direction of the spectator quark in the top-quark (top-antiquark) reference frame.
The three components of
polarisation are measured to be $\Polx =
0.01\pm0.18$, $\Poly = -0.029\pm0.027$, $\Polz =
0.91\pm0.10$ for top quarks and $\Polx = -0.02\pm0.20$, $\Poly = -0.007\pm0.051$,
$\Polz = -0.79\pm 0.16$ for top antiquarks.
The results are consistent with NNLO QCD predictions and expectation of $\Poly^{t}=\Poly^{\bar t}=0$ from
the hypothesis of CP symmetry in the top-quark and top-antiquark decay.

Normalised differential cross-sections for top-quark production are measured at particle level
as a function of the emission angle of the charged lepton resulting from the
$t \rightarrow Wb \rightarrow \ell\nu b$ decay. They are provided for top quarks and antiquarks, both inclusively and separately.
Such distributions are associated with the top-quark polarisation components.
The measurements are consistent with SM predictions provided by
various MC generators at LO and NLO in QCD.
 
An EFT prediction is fitted to the measured differential cross-sections to obtain
exclusion limits simultaneously for the real and imaginary parts of the $\mathcal{O}_{tW}$ operator.
Using a morphing technique, 95\%~CL intervals for these operators are found to be $\ctW \in [-0.9,1.4]$ and $\citW \in [-0.8,0.2]$,
compatible with the SM predictions obtained with the NLO $\MGNLOPY8$ generator.
The bounds on \citW\ presented in this paper improve on previous results from ATLAS.

\section*{Acknowledgements}


We thank CERN for the very successful operation of the LHC, as well as the
support staff from our institutions without whom ATLAS could not be
operated efficiently.
 
We acknowledge the support of
ANPCyT, Argentina;
YerPhI, Armenia;
ARC, Australia;
BMWFW and FWF, Austria;
ANAS, Azerbaijan;
SSTC, Belarus;
CNPq and FAPESP, Brazil;
NSERC, NRC and CFI, Canada;
CERN;
ANID, Chile;
CAS, MOST and NSFC, China;
Minciencias, Colombia;
MEYS CR, Czech Republic;
DNRF and DNSRC, Denmark;
IN2P3-CNRS and CEA-DRF/IRFU, France;
SRNSFG, Georgia;
BMBF, HGF and MPG, Germany;
GSRI, Greece;
RGC and Hong Kong SAR, China;
ISF and Benoziyo Center, Israel;
INFN, Italy;
MEXT and JSPS, Japan;
CNRST, Morocco;
NWO, Netherlands;
RCN, Norway;
MEiN, Poland;
FCT, Portugal;
MNE/IFA, Romania;
JINR;
MES of Russia and NRC KI, Russian Federation;
MESTD, Serbia;
MSSR, Slovakia;
ARRS and MIZ\v{S}, Slovenia;
DSI/NRF, South Africa;
MICINN, Spain;
SRC and Wallenberg Foundation, Sweden;
SERI, SNSF and Cantons of Bern and Geneva, Switzerland;
MOST, Taiwan;
TENMAK, T\"urkiye;
STFC, United Kingdom;
DOE and NSF, United States of America.
In addition, individual groups and members have received support from
BCKDF, CANARIE, Compute Canada and CRC, Canada;
COST, ERC, ERDF, Horizon 2020 and Marie Sk{\l}odowska-Curie Actions, European Union;
Investissements d'Avenir Labex, Investissements d'Avenir Idex and ANR, France;
DFG and AvH Foundation, Germany;
Herakleitos, Thales and Aristeia programmes co-financed by EU-ESF and the Greek NSRF, Greece;
BSF-NSF and GIF, Israel;
Norwegian Financial Mechanism 2014-2021, Norway;
NCN and NAWA, Poland;
La Caixa Banking Foundation, CERCA Programme Generalitat de Catalunya and PROMETEO and GenT Programmes Generalitat Valenciana, Spain;
G\"{o}ran Gustafssons Stiftelse, Sweden;
The Royal Society and Leverhulme Trust, United Kingdom.
 
The crucial computing support from all WLCG partners is acknowledged gratefully, in particular from CERN, the ATLAS Tier-1 facilities at TRIUMF (Canada), NDGF (Denmark, Norway, Sweden), CC-IN2P3 (France), KIT/GridKA (Germany), INFN-CNAF (Italy), NL-T1 (Netherlands), PIC (Spain), ASGC (Taiwan), RAL (UK) and BNL (USA), the Tier-2 facilities worldwide and large non-WLCG resource providers. Major contributors of computing resources are listed in Ref.~\cite{ATL-SOFT-PUB-2021-003}.
 

\printbibliography
 
\clearpage 
 
\begin{flushleft}
\hypersetup{urlcolor=black}
{\Large The ATLAS Collaboration}

\bigskip

\AtlasOrcid[0000-0002-6665-4934]{G.~Aad}$^\textrm{\scriptsize 99}$,    
\AtlasOrcid[0000-0002-5888-2734]{B.~Abbott}$^\textrm{\scriptsize 124}$,    
\AtlasOrcid[0000-0002-7248-3203]{D.C.~Abbott}$^\textrm{\scriptsize 100}$,    
\AtlasOrcid[0000-0002-2788-3822]{A.~Abed~Abud}$^\textrm{\scriptsize 34}$,    
\AtlasOrcid[0000-0002-1002-1652]{K.~Abeling}$^\textrm{\scriptsize 51}$,    
\AtlasOrcid[0000-0002-2987-4006]{D.K.~Abhayasinghe}$^\textrm{\scriptsize 91}$,    
\AtlasOrcid[0000-0002-8496-9294]{S.H.~Abidi}$^\textrm{\scriptsize 27}$,    
\AtlasOrcid[0000-0001-5329-6640]{H.~Abramowicz}$^\textrm{\scriptsize 157}$,    
\AtlasOrcid[0000-0002-1599-2896]{H.~Abreu}$^\textrm{\scriptsize 156}$,    
\AtlasOrcid[0000-0003-0403-3697]{Y.~Abulaiti}$^\textrm{\scriptsize 5}$
% ...
\bigskip
\\

% ...
$^{5}$High Energy Physics Division, Argonne National Laboratory, Argonne IL; United States of America.\\
% ...
$^{27}$Physics Department, Brookhaven National Laboratory, Upton NY; United States of America.\\
% ...
$^{34}$CERN, Geneva; Switzerland.\\
% ...
$^{51}$II. Physikalisches Institut, Georg-August-Universit\"{a}t G\"ottingen, G\"ottingen; Germany.\\
% ...
$^{91}$Department of Physics, Royal Holloway University of London, Egham; United Kingdom.\\
% ...
$^{99}$CPPM, Aix-Marseille Universit\'e, CNRS/IN2P3, Marseille; France.\\
$^{100}$Department of Physics, University of Massachusetts, Amherst MA; United States of America.\\
% ...
$^{124}$Homer L. Dodge Department of Physics and Astronomy, University of Oklahoma, Norman OK; United States of America.\\
% ...
$^{156}$Department of Physics, Technion, Israel Institute of Technology, Haifa; Israel.\\
$^{157}$Raymond and Beverly Sackler School of Physics and Astronomy, Tel Aviv University, Tel Aviv; Israel.\\
% ...

\end{flushleft}


\end{document}